\documentclass[aps,12pt,floatfix,longbibliography,a4paper]{revtex4-2}
\usepackage{graphicx}
\usepackage{epstopdf}
\usepackage{amssymb}
\usepackage{hyperref}
\usepackage[normalem]{ulem}
\usepackage{color}
\usepackage{amsmath,bm}
\usepackage[utf8]{inputenc}
\usepackage{xcolor}
\usepackage{physics}
\usepackage{mathcomp}
\usepackage{multirow}
\usepackage{comment}
\usepackage{makecell}
\usepackage{longtable}

\usepackage[labeled]{multibib}
\usepackage{placeins}

\newcommand{\LNO}{La$_3$Ni$_2$O$_7$}
\newcommand{\LPNO}{La$_2$PrNi$_2$O$_7$}
\newcommand{\LMNO}{La$_2$SmNi$_2$O$_7$}

\newcounter{hoge}
\makeatletter
\@addtoreset{figure}{hoge}%%%(*2)
\makeatother
\makeatletter
\@addtoreset{equation}{hoge}%%%(*2)
\makeatother
\makeatletter
\@addtoreset{section}{hoge}%%%(*2)
\makeatother
\makeatletter
\@addtoreset{table}{hoge}%%%(*2)
\makeatother

\begin{document}

\title[]{Polar, checkerboard charge order in bilayer nickelate \LNO{}}

%%=============================================================%%

\author{Ryo Misawa$^{1}$}
\email{misawann6@g.ecc.u-tokyo.ac.jp}
\author{Shunsuke Kitou$^{2}$}
\author{Jian-Ping Sun$^{3,4}$}
\author{Yingpeng Yu$^{3,4}$}
\author{Chihaya Koyama$^{2}$}
\author{Lin Zhu$^{3,4}$}
\author{Ningning Wang$^{5}$}
\author{Yuiga Nakamura$^{6}$}
\author{Kouhei Ichiyanagi$^{6}$}
\author{Toshiyuki Sasaki$^{6}$}
\author{Taka-hisa Arima$^{2,7}$}
\author{Jin-Guang Cheng$^{3,4}$}
\author{Max Hirschberger$^{1}$}
\email{hirschberger@ap.t.u-tokyo.ac.jp}

\affiliation{$^{1}$Department of Applied Physics, The University of Tokyo, Bunkyo, Tokyo 113-8656, Japan}
\affiliation{$^{2}$Department of Advanced Materials Science, The University of Tokyo, Kashiwa, Chiba 277-8561, Japan}
\affiliation{$^{3}$Beijing National Laboratory for Condensed Matter Physics, and Institute of Physics, Chinese Academy of Sciences, Beijing 100190, China}
\affiliation{$^{4}$School of Physical Sciences, University of Chinese Academy of Sciences, Beijing 100190, China}
\affiliation{$^{5}$Institute of Energy, Hefei Comprehensive National Science Center (Anhui Energy Laboratory), Hefei 230031, China}
\affiliation{$^{6}$Japan Synchrotron Radiation Research Institute (JASRI), SPring-8, Hyogo 679-5198, Japan}
\affiliation{$^{7}$RIKEN Center for Emergent Matter Science (CEMS), Wako, Saitama 351-0198, Japan}

\maketitle
\newpage

\begin{center}
\Large{Abstract}
\end{center}

\textbf{We revisit the crystal structure of bilayer nickelates \LNO{} and \LPNO{} to reveal their charge order under ambient conditions. Using synchrotron X-ray diffraction on high-quality single crystals, we resolve faint reflections---nearly four orders of magnitude weaker than the main Bragg reflections in \LNO{} and an additional order weaker in \LPNO{}---that were overlooked in prior diffraction studies. These reflections evidence a polar crystal structure, rather than the widely used centrosymmetric structure model. The polarity is induced by checkerboard charge order in combination with octahedral tilting. Furthermore, we directly observe a melting of the charge order above $T_\mathrm{CO}=560\,$K, accompanied by an orthorhombic-to-orthorhombic structural phase transition. This work proposes the checkerboard charge order as a precursor to the spin density wave order at a lower temperature and as a competing phase of the superconducting state in bilayer nickelates.}

\newpage

\begin{center}
\Large{Main Text}
\end{center}
%%%%%%%%%%%%%%%%%%%%%%%%%%%%%
%%% SECTION
%%%%%%%%%%%%%%%%%%%%%%%%%%%%%
Transition metal oxides with the Ruddlesden-Popper (RP) structure constitute a prominent class of strongly correlated quantum materials, offering versatile functionalities for electronic and magnonic applications. The RP oxides are characterized by the chemical formula of $A_{n+1}$$B_{n}$O$_{3n+1}$ with $n$ layers of the perovskite $AB$O$_3$ block. One of the most typical systems is the bilayer phase $A_3$$B_2$O$_7$. Across a wide range of transition metals, including manganese, iron, cobalt, and ruthenium, the bilayer RP materials host strongly correlated phenomena, such as colossal magnetoresistance~\cite{Moritomo1996-vz}, ferroelectricity~\cite{Lee2013-tj,Oh2015-vd,Smith2019-nn}, charge ordering~\cite{Tokunaga2006-bz,Kim2021-vw,Huang2026-ax}, and polar metallicity~\cite{Lei2018-lg,Zhou2026-uk}. Despite their centrosymmetric parent structures, polarity in ferroelectrics and metals can arise from the combination of multiple octahedral distortions~\cite{Benedek2011-kv,Oh2015-vd,Lei2018-lg}, or from charge order~\cite{Tokunaga2006-bz,Zhou2026-uk,Huang2026-ax}. The resulting broken-inversion symmetry, when coupled to magnetism, facilitates multiferroic responses~\cite{Benedek2011-kv,Zhou2026-uk,Liu2018-jx}.

\begin{figure}[ht]%
\centering
\includegraphics[width=1.0\textwidth]{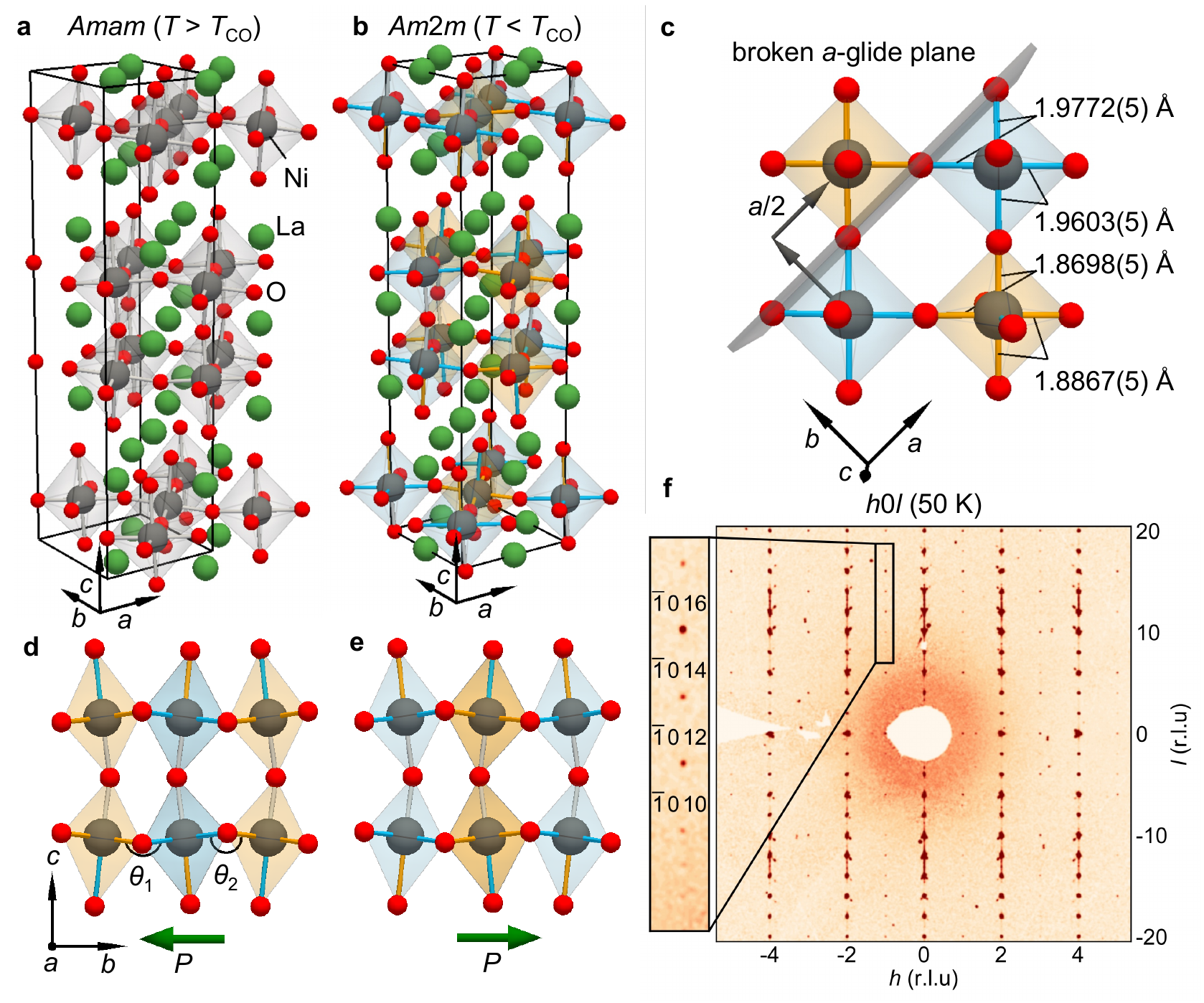}
\caption{\textbf{Polar, checkerboard charge order in bilayer nickelate \LNO{} at ambient pressure.}
\textbf{a}, Centrosymmetric, orthorhombic $Amam$ structure model of \LNO{}, previously reported in the literature under ambient conditions, and identified in the present study as a high-temperature phase above $T_\mathrm{CO} = 560\,$K (CO = charge order). There is only a single independent nickel site, surrounded by gray oxygen octahedra. \textbf{b}, Polar orthorhombic $Am2m$ structure proposed in the present work. This model contains two inequivalent nickel sites with contracted (orange) and expanded (blue) NiO$_6$ octahedral environments. The orange and blue lines indicate shorter and longer Ni-O bonds. \textbf{c}, Broken $a$-glide symmetry in the $Am2m$ structure. The glide operation maps a nickel site to the other independent site; the symmetry is broken if the sites are inequivalent. Broken glide plane indicated in gray. \textbf{d}, Polarity $P$ induced by the charge alternation along the $b$ axis in combination with the bond alternation due to octahedral tilting around the $a$ axis. The orange and blue octahedra carry a bond-valence sum of $+2.7852(15)$ and $+2.3466(12)$, respectively (see Methods). The tilting angles are $\theta_1 = 192.183(16)^\circ$ and $\theta_2=170.945(14)^\circ$. \textbf{e}, Another polar domain obtained by swapping charges in panel~\textbf{d}. \textbf{f}, X-ray diffraction pattern on the $h0l$ plane at $50~$K. Here, $h0l$ ($h=-1$) reflections associated with the broken $a$-glide plane are highlighted by the black rectangle.
}
\label{main:Fig1}
\end{figure}

More recently, the discovery of high-temperature superconductivity under pressure in both bilayer and trilayer RP nickelates has opened a new frontier for this materials family~\cite{Sun2023-sf,Zhang2024-uu,Zhu2024-rn}. Focusing on the bilayer nickelate \LNO{}, a majority of crystals are reported to crystallize in the orthorhombic $Amam$ structure at ambient pressure (Fig.~\ref{main:Fig1}\textbf{a}). Distinct structures built from monolayer and trilayer blocks are known polymorphs of \LNO{} (see Methods). The widely used $Amam$ structure is inferred largely from powder X-ray diffraction (XRD)~\cite{Zhang1994-vm,Taniguchi1995-lg,Ling2000-uz,Voronin2001-oc,Wang2024-qi} and from in-house single-crystal XRD~\cite{Wang2025-tg,Chen2024-dr,Wang2024-ii}. This structure model contains a single nickel site with the nominal valence of $2.5+$. The same model is employed for cation-engineered systems, such as \LPNO{} and \LMNO{}, where enhanced superconductivity is reported, possibly through cation-induced changes in the lattice geometry and oxygen stoichiometry~\cite{Wang2024-dq,Dong2025-ni,Li2026-go}. The mixed-valence state of nickel suggests a potential for bilayer nickelates to show charge order at ambient pressure, thus highlighting the need to clarify the structural properties with high experimental resolution. Here, we demonstrate polar charge order in \LNO{} and \LPNO{} at ambient conditions (Fig.~\ref{main:Fig1}\textbf{b}-\textbf{e}), directly evidenced by a broken glide symmetry in synchrotron XRD experiments.

%%%%%%%%%%%%%%%%%%%%%%%%%%%%%
%%% SECTION
%%%%%%%%%%%%%%%%%%%%%%%%%%%%%
\bigskip
\textbf{Breaking of glide symmetry}

To precisely investigate the crystal structure of \LNO{} at ambient pressure, we perform synchrotron X-ray diffraction on a high-quality single crystal in the bending-magnet beamline BL02B1 (Methods). We use an area detector with a dynamic range of $10^6$, enabling simultaneous detection of weak oxygen-derived intensities and much stronger reflections dominated by lanthanum and nickel. From systematic extinction of reflections in the observed diffraction patterns, we identify an orthorhombic $A$-base-centered lattice at and below room temperature, consistent with previous reports~\cite{Zhang1994-vm,Taniguchi1995-lg,Ling2000-uz,Voronin2001-oc,Wang2024-qi,Wang2025-tg}. However, there is evidence of symmetry breaking in the plane of reciprocal space described by the Miller indices $h0l$, which was not previously reported. Figure~\ref{main:Fig1}\textbf{f} shows the XRD diffraction pattern on this plane, taken at temperature $T=50\,$K. We resolve the systematic presence of $h0l$ ($h=$~odd) reflections. This corresponds to the absence of the $a$-glide plane, in clear disagreement with the reported space group $Amam$~\cite{Zhang1994-vm,Taniguchi1995-lg,Ling2000-uz,Voronin2001-oc,Wang2024-qi,Wang2025-tg}. Since these reflections are observed at and below room temperature, we hereafter focus on the data at $T=50\,$K, at which the intensities are the strongest among the measured temperatures.

\begin{figure}[ht]%
\centering
\includegraphics[width=0.5\textwidth]{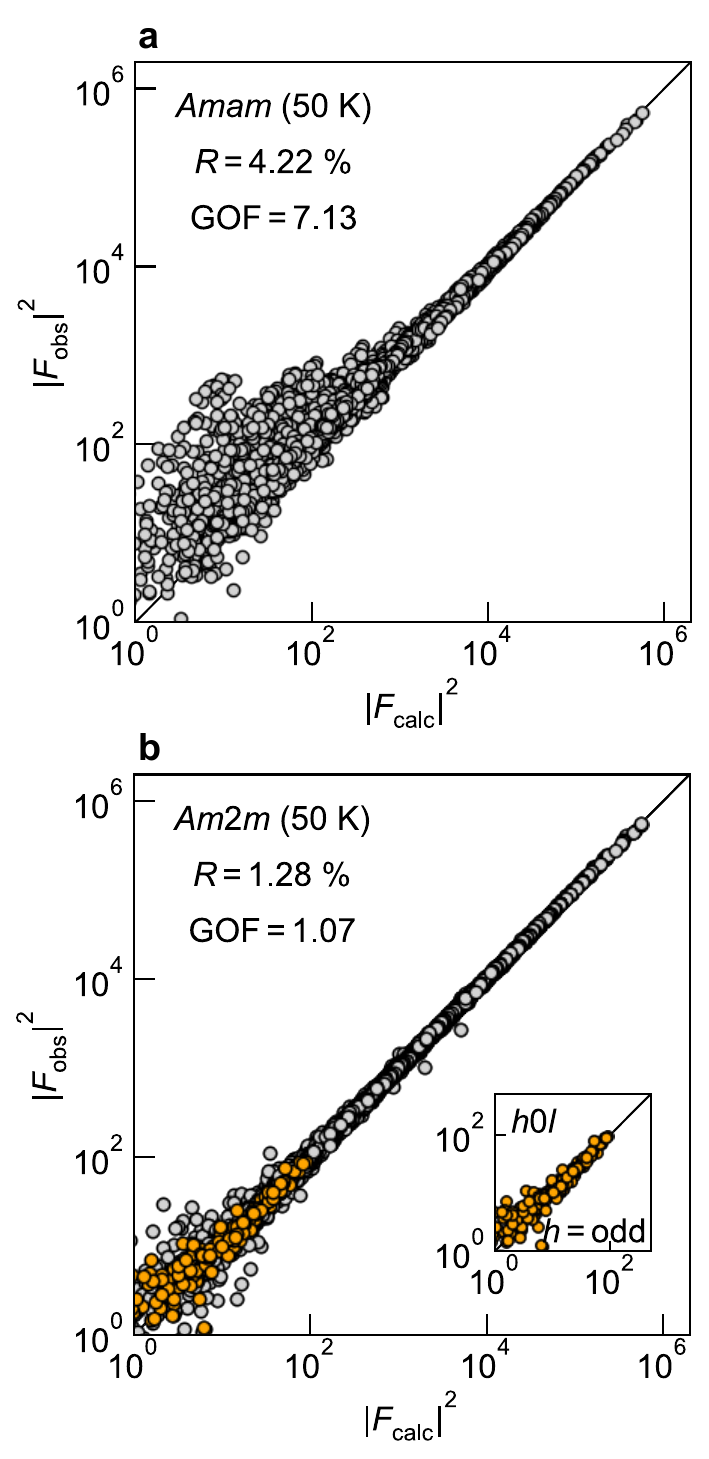}
\caption{\textbf{Structure refinement based on synchrotron X-ray diffraction (XRD).}
\textbf{a}, Refinement of XRD data at $T = 50\,$K assuming the $Amam$ structure model. Observed and calculated intensities, expressed in terms of $|F_\mathrm{obs}|^2$ and $|F_\mathrm{calc}|^2$, show a poor agreement in the regime of weak intensity. Here, $F_\mathrm{obs}$ and $F_\mathrm{calc}$ denote the observed and calculated structure factors, respectively. The poor agreement of model and experiment is quantified by the reliability factor $R = 4.22\,\%$ and goodness-of-fit (GOF) of $7.13 \gg 1$. \textbf{b}, Refinement assuming the proposed $Am2m$ structure. $|F_\mathrm{obs}|^2$ and $|F_\mathrm{calc}|^2$ show a much better agreement with $R = 1.28\,\%$ and $\text{GOF}=1.07$. Inset focuses on $h0l$ ($h=\text{odd}$) reflections (orange circles) associated with the broken $a$-glide symmetry.
}
\label{main:Fig2}
\end{figure}

Quantitative evidence against the $Amam$ structure is provided by high-precision structure refinement, enabled by an excellent coverage of reciprocal space: $98\,\%$ of all available reflections, up to a resolution of $0.3249\,$\AA, were measured by rotating the sample in fine steps, yielding the redundancy $16.296$ (Methods). Figure~\ref{main:Fig2}\textbf{a} compares the observed and calculated XRD intensities for the best refinement based on the $Amam$ structure, in terms of  $|F_\mathrm{obs}|^2$ and $|F_\mathrm{calc}|^2$. Here, $F_\mathrm{obs}$ and $F_\mathrm{calc}$ represent the observed and calculated structure factors, respectively. Even when ignoring the reflections that break the $a$-glide plane as symmetry-extinct in the refinement, the result shows a poorer agreement between $|F_\mathrm{obs}|^2$ and $|F_\mathrm{calc}|^2$ with reliability factor $R = 4.22\,\%$ and goodness-of-fit (GOF) of $7.13$ based on $5,542$ unique reflections; see Table~\ref{Tab:refinement} for other figures of merit of this refinement. The thermal ellipsoids of oxygen atoms show an anomalously flat shape in the refinement using $Amam$, further suggesting the need to lower the structural symmetry~(Supplementary Fig.~\ref{SI:thermal}).

To identify the correct space group, we consider subgroups of $Amam$; we find that $Amam$ does exist in bilayer nickelates at a much higher temperature, as demonstrated later. The reflection conditions uniquely determine the highest-symmetry space group candidate as polar $Am2m$. Indeed, our refinement of XRD intensities using this structure yields much better agreement, with $R=1.28\,\%$ and $\text{GOF}=1.07$ based on $10,834$ unique reflections (Fig.~\ref{main:Fig2}\textbf{b}). The structure refinement further reveals unequal, though far from single-domain, populations of the polar domains, with fractions of $0.63(2)$ and $0.37(2)$, as determined from the Flack parameter. Moreover, the refined oxygen occupancies show no deviation from the stoichiometric value, and thermal ellipsoids at the oxygen sites are small and regular in shape (see Supplementary Note~\ref{sec:SI_str}). These results indicate that the faint streaks in the XRD pattern (Fig.~\ref{main:Fig1}\textbf{f}), much weaker than the Bragg reflections, are consistent with minor stacking faults, in contrast to the severe diffuse streaks arising from stacking disorder in a polymorph of \LNO{} (see Methods). Together, our analysis excludes the widely assumed $Amam$ structure and determines the proper structure as $Am2m$.

%%%%%%%%%%%%%%%%%%%%%%%%%%%%%
%%% SECTION
%%%%%%%%%%%%%%%%%%%%%%%%%%%%%
\bigskip
\textbf{Discovery of polar, checkerboard charge order in \LNO{}}

We now examine the refined structure in detail. Unlike the $I4/mmm$ and $Amam$ structure models, the proposed $Am2m$ structure contains two inequivalent nickel sites, as shown in Fig~\ref{main:Fig1}\textbf{b},\textbf{c}. These two sites have distinct Ni-O bond lengths, whose difference is rather large, $\sim 0.1~\text{\AA}$. We note that a recent report lowered symmetry to monoclinic $P2_1/m$ based on in-house single-crystal XRD, but found Ni–O bond-length differences of only $0.01$~\AA\ between inequivalent Ni sites~\cite{Li2026-go}. This is an order of magnitude smaller than in our synchrotron XRD refinement (Fig.~\ref{main:Fig1}\textbf{c}); it may be difficult to resolve weak intensities related to oxygen displacements in in-house XRD, without the high synchrotron photon flux and large dynamic range employed in our experiment. The reflection conditions in our data also exclude the $P2_1/m$ structure (Supplementary Note~\ref{sec:monocli}).

The two distinct octahedral volumes with the large difference in the Ni-O bond lengths are consistent with charge order, a reasonable scenario for an oxide with a nominal nickel valence of $2.5+$. From the well-established bond-valence sum (BVS) approach (Methods), we estimate valences of $+2.7852(15)$ and $+2.3466(12)$ for the contracted and expanded octahedra, respectively. We note that the BVS does not distinguish whether the charge order resides primarily on the nickel sites or on the ligand oxygen. The absence of the $a$-glide symmetry is depicted in Fig.~\ref{main:Fig1}\textbf{c}, where the glide operation connects the two nickel sites with different ligand volumes. Here, the two nickel sites alternate in each layer and form a checkerboard pattern of charge. This checkerboard charge order, combined with octahedral tilting of the oxygen ligand environment, imparts the polarity to \LNO{}. Figure~\ref{main:Fig1}\textbf{d} illustrates a single polar domain as seen from the side; its inversion pair is obtained by swapping the two nickel sites (Fig.~\ref{main:Fig1}\textbf{e}). Note that, without octahedral tilting, the checkerboard charge distribution within a single layer would result in full cancellation of the net polarity. Only when octahedral tilting introduces bond alternation along the same direction ($b$ axis) as the site alternation, cancellation no longer occurs, and a macroscopic polarity arises.

%%%%%%%%%%%%%%%%%%%%%%%%%%%%%
%%% SECTION
%%%%%%%%%%%%%%%%%%%%%%%%%%%%%
\bigskip
\textbf{High-temperature charge-ordering transition}

\begin{figure}[ht]%
\centering
\includegraphics[width=0.9\textwidth]{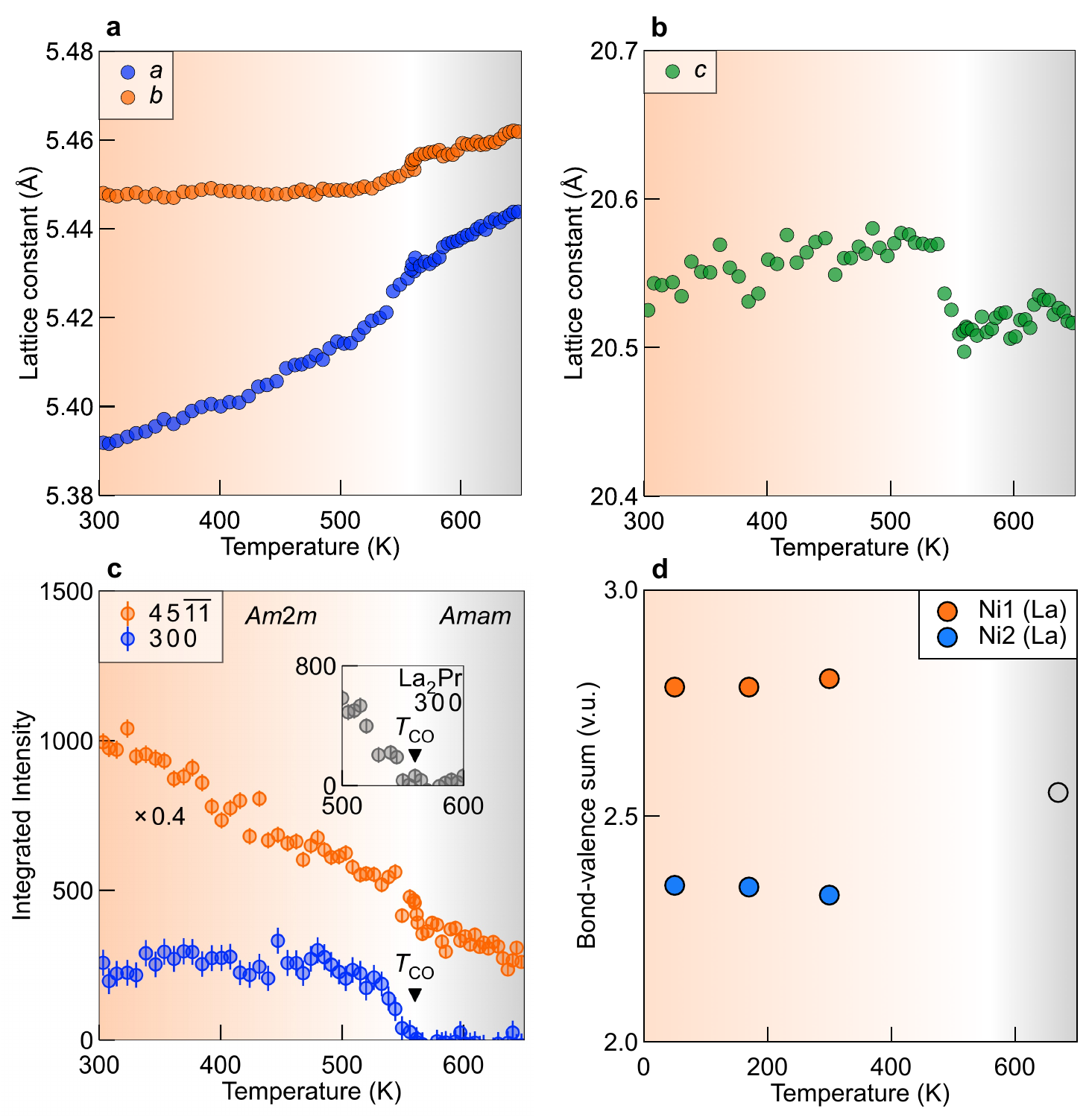}
\caption{\textbf{Direct observation of the charge-ordering transition in \LNO{}.}
\textbf{a}, Temperature dependence of the lattice constants $a$ (blue) and $b$ (orange). These values show an anomaly at $T_\mathrm{CO}=560\,$K (CO: charge order) and get close to each other above $T_\mathrm{CO}$. The residual difference between $a$ and $b$ indicates an orthorhombic structure above $T_\mathrm{CO}$. \textbf{b}, Temperature dependence of the lattice constant $c$. Just below $T_\mathrm{CO}$, the value of $c$ anomalously increases. \textbf{c}, Temperature dependence of X-ray diffraction intensities. Above $T_\mathrm{CO}$, the intensity of $4\,5\,\overline{1}$ (orange) remains, while it should vanish for a tetragonal $I4/mmm$ structure. On the other hand, $3\,0\,0$ (blue) disappears above $T_\mathrm{CO}$, evidencing a melting of charge order that accompanies a structural transition from $Am2m$ to $Amam$. Inset: Temperature dependence of the intensity at $3\,0\,0$ for \LPNO{}. \textbf{d}, Temperature dependence of the bond-valence sum (BVS) for NiO$_6$. In the charge-ordered state, the BVSs are nearly temperature-independent at $2.8\,$v.u. (valence units, orange) and $2.3\,$v.u. (blue). Above $T_\mathrm{CO}$, the BVS is merged to $2.6\,$v.u., consistent with the mixed valence of nickel in $Amam$. In panel~\textbf{d}, the data points below $T_\mathrm{CO}$ are from piece P$1$, while the one above $T_\mathrm{CO}$ is from P$2$.
}
\label{main:Fig3}
\end{figure}

To observe the melting of charge order directly, we perform synchrotron XRD experiments at higher temperatures. We employ a twinned single crystal (piece P2), rather than the one with a single orthorhombic domain used for the low-temperature measurements (P1). This enables us to evaluate the temperature evolution of orthorhombic twins; if the high-temperature phase were tetragonal $I4/mmm$, peak splitting related to rotation symmetry breaking in the basal plane would completely disappear. Figure~\ref{main:Fig3}\textbf{a} illustrates the temperature dependence of the lattice constants $a$ and $b$. Both exhibit an anomaly at $T_\mathrm{CO} = 560\,$K (CO = charge order). As evidenced by the decreasing separation between the split Bragg peaks (Supplementary Fig.~\ref{SI:domain}), $a$ and $b$ become closer as temperature increases towards $T_\mathrm{CO}$. However, their difference remains finite even above $T_\mathrm{CO}$, demonstrating the persistence of an orthorhombic structure. Also, the lattice constant $c$ exhibits an anomalous decrease just above $T_\mathrm{CO}$, as shown in Fig.~\ref{main:Fig3}\textbf{b}. A prior study identifies a similar temperature dependence of the lattice constants with the anomalous behavior around $T_\mathrm{CO}$ in oxygen-deficient La$_3$Ni$_2$O$_{6.92}$~\cite{Sasaki1997-qj}; this suggests the existence of this phase transition even in non-stoichiometric crystals.

To identify the crystal structure above $T_\mathrm{CO}$, we illustrate the temperature dependence of XRD intensities in Fig.~\ref{main:Fig3}\textbf{c}. Here, $4\, 5\, \overline{1}$ corresponds to a forbidden reflection in the parent tetragonal $I4/mmm$ structure, whereas $3\, 0\, 0$ is related to the presence of the $a$-glide plane; in other words, an order parameter of the charge order. The $4\, 5\, \overline{1}$ reflection remains even above $T_\mathrm{CO}$, and this indicates that the crystal lattice is still orthorhombic $A$-base-centered, consistent with the residual difference between $a$ and $b$ (Fig.~\ref{main:Fig3}\textbf{a}). On the other hand, $3\, 0\, 0$ disappears above $T_\mathrm{CO}=560\,$K. This directly demonstrates the melting of charge order at $T_\mathrm{CO}$, accompanied by a structural phase transition from orthorhombic $Am2m$ to orthorhombic $Amam$, as further confirmed by structure refinement (Supplementary Note~\ref{sec:SI_str}). Figure~\ref{main:Fig3}\textbf{d} illustrates the temperature dependence of the BVS. In the charge-ordered state, BVS is split into two values, which are nearly independent of temperature at $2.8+$ and $2.3+$. Above $T_\mathrm{CO}$, he valences merge into a single value of $+2.552(15)$, in line with the mixed valence of nickel in $Amam$.\\

\textbf{Weaker charge order in \LPNO{}}

To investigate the impact of cation engineering on charge order, we performed the XRD measurements in the same bending-magnet BL02B1 beamline. Due to the positive chemical pressure by praseodymium substitution, the unit cell volume $V=598.960(18)\,\mathrm{\AA}^3$ is smaller than $V=603.534(18)\,\mathrm{\AA}^3$ in \LNO{} at room temperature. However, we did not observe the reflections associated with broken $a$-glide symmetry at this beamline.

We therefore used the BL05XU undulator beamline, which provides synchrotron X-rays with an intensity approximately two to three orders of magnitude higher than that of BL02B1. This enables us to observe $h\,0\,l$ ($h=\,$odd) reflections (Supplementary Fig.~\ref{SI:Pr}\textbf{a}). These reflections related to charge order are five orders of magnitude weaker than the main Bragg reflections, i.e., ten times weaker than in \LNO{}. Structure refinement using the $Am2m$ model shows that praseodymium preferentially occupies the outer rock-salt layer (Supplementary Fig.~\ref{SI:Pr}\textbf{b}), which agrees with previous electron energy loss spectroscopy~\cite{Dong2025-ni}. The difference between the BVS at two nickel sites is weaker than that for \LNO{}, consistent with the weaker signals of $a$-glide symmetry breaking. We further followed the temperature dependence of the $3\,0\,0$ reflection, observing its emergence below $T_\mathrm{CO}=560\,$K (Fig.~\ref{main:Fig3}\textbf{c}). The charge-ordering transition temperature thus remains almost unchanged upon praseodymium substitution.

%%%%%%%%%%%%%%%%%%%%%%%%%%%%%
%%% SECTION
%%%%%%%%%%%%%%%%%%%%%%%%%%%%%
\bigskip
\textbf{Discussion and Conclusion}

A reasonable microscopic description of the charge order is ligand-hole states, as suggested by spectroscopic studies~\cite{Chen2024-rixs,Takegami2024-fs}. This is in contrast to charge order exclusively on the nickel sites. Such charge order is related to the phenomenology of bond disproportionation and has been proposed and identified in perovskite \textit{R}NiO$_3$ ($R=$ rare-earth)~\cite{Johnston2014-jt,Green2016-du,Lu2018-fd}. The concomitant Zhang-Rice singlet formation between the holes on the ligand and nickel, proposed theoretically~\cite{Jiang2025-bu,Foyevtsova2025-fs}, may account for the spin density wave order in \LNO{} below $150\,$K, with two distinct moment sizes (Supplementary Note~\ref{sec:ZRS}). Indeed, a prior theoretical study simultaneously stabilizes checkerboard charge and spin orders~\cite{LaBollita2024-rd}.

The checkerboard charge order discovered here is a plausible yet previously unrecognized state in the bilayer nickel oxide \LNO{} with mixed valence of nickel. Indeed, our charge order harks back to the structure of the bilayer manganese oxide Pr(Sr$_{1-x}$Ca$_x$)$_2$Mn$_2$O$_7$, which likewise exhibits electric polarization due to the alternation of sites and bonds~\cite{Tokunaga2006-bz}. Related, yet different, structures have been proposed in several bilayer Ruddlesden-Popper transition-metal oxides: out-of-plane polarity of Sr$_3$Co$_2$O$_7$ is theoretically linked to charge order~\cite{Zhou2026-uk,Huang2026-ax}, in-plane polarity in Ca$_3$Ru$_2$O$_7$ arises from tilts and rotations of octahedra~\cite{Benedek2011-kv,Lei2018-lg}, and hidden order in Sr$_3$Fe$_2$O$_7$ is ascribed to centrosymmetric checkerboard charge order by resonant X-ray diffraction~\cite{Kim2021-vw}. As in these transition-metal oxides, our work on \LNO{} anticipates electrical, optical, and magnetoelectric functionalities arising from the coupling between the polar charge order and low-temperature magnetic order~\cite{Tokunaga2006-bz,Zhou2026-uk}.

The concurrence of superconductivity and a structural phase transition under pressure~\cite{Sun2023-sf,Wang2024-qi} suggests that the ambient-pressure charge order suppresses superconductivity. Further, cation substitution may introduce local disorder to disturb and weaken charge order within the bilayer. Future comprehensive studies need to clarify the interplay among cation engineering, charge order, and superconductivity through single-crystal XRD under pressure. Together, this work reshapes the current understanding of the Ruddlesden-Popper nickel oxides and motivates further experimental and theoretical studies of emergent responses from the polar charge order and its interplay with magnetism and superconductivity.

\textbf{Note added:} A subsequent study confirms the checkerboard charge order in a stoichiometric \LNO{} single crystal, grown by a different method~\cite{Liu2026-vk}. The proposed $Am2m$ structure has been shown to account for previously unexplained features in the $^{139}$La nuclear quadrupole resonance (NQR) spectra of polycrystalline \LNO{} prepared by solid-state reaction~\cite{Yashima2026-qv}.

\newpage
\bibliography{sn-bibliography}

@ARTICLE{Petricek2014-pc,
  title     = "{Crystallographic Computing System JANA2006: General features}",
  author    = "Petříček, Václav and Dušek, Michal and Palatinus, Lukáš",
  journal   = "Z. Kristallogr. Cryst. Mater.",
  publisher = "Walter de Gruyter GmbH",
  volume    =  229,
  number    =  5,
  pages     = "345--352",
  month     =  "1~" # may,
  year      =  2014
}

@MANUAL{Agilent-Technologies-Ltd2014-hl,
  title        = "{CrysAlisPro}",
  author       = "{Agilent Technologies Ltd}",
  address      = "Yarnton, Oxfordshire, England",
  year         =  2014,
  organization = "Agilent Technologies Ltd"
}

@ARTICLE{Tokunaga2006-bz,
  title     = "{{Rotation of orbital stripes and the consequent charge-polarized
               state in bilayer manganites}}",
  author    = "Tokunaga, Yusuke and Lottermoser, Thomas and Lee, Yunsang and
               Kumai, Reiji and Uchida, Masaya and Arima, Takahisa and Tokura,
               Yoshinori",
  journal   = "Nat. Mater.",
  publisher = "Springer Science and Business Media LLC",
  volume    =  5,
  number    =  12,
  pages     = "937--941",
  month     =  "5~" # dec,
  year      =  2006
}

@ARTICLE{Sun2023-sf,
  title     = "{{Signatures of superconductivity near $80$~K in a nickelate under high pressure}}",
  author    = "Sun, Hualei and Huo, Mengwu and Hu, Xunwu and Li, Jingyuan and
               Liu, Zengjia and Han, Yifeng and Tang, Lingyun and Mao, Zhongquan
               and Yang, Pengtao and Wang, Bosen and Cheng, Jinguang and Yao,
               Dao-Xin and Zhang, Guang-Ming and Wang, Meng",
  journal   = "Nature",
  publisher = "Nature Publishing Group",
  volume    =  621,
  number    =  7979,
  pages     = "493--498",
  month     =  "12~" # sep,
  year      =  2023
}

@ARTICLE{Zhang2024-uu,
  title     = "{{High-temperature superconductivity with zero resistance and
               strange-metal behaviour in {La$_3$Ni$_2$O$_{7-\delta}$}}}",
  author    = "Zhang, Yanan and Su, Dajun and Huang, Yanen and Shan, Zhaoyang
               and Sun, Hualei and Huo, Mengwu and Ye, Kaixin and Zhang, Jiawen
               and Yang, Zihan and Xu, Yongkang and Su, Yi and Li, Rui and
               Smidman, Michael and Wang, Meng and Jiao, Lin and Yuan, Huiqiu",
  journal   = "Nat. Phys.",
  publisher = "Springer Science and Business Media LLC",
  volume    =  20,
  number    =  8,
  pages     = "1269--1273",
  month     =  "6~" # aug,
  year      =  2024
}

@ARTICLE{Taniguchi1995-lg,
  title     = "{{Transport, magnetic and thermal properties of {La}$_{3}${Ni}$_2${O}$_{7-\delta}$}}",
  author    = "Taniguchi, Satoshi and Nishikawa, Takashi and Yasui, Yukio and
               Kobayashi, Yoshiaki and Takeda, Jun and Shamoto, Shin-Ichi and
               Sato, Masatoshi",
  journal   = "J. Phys. Soc. Jpn.",
  publisher = "Physical Society of Japan",
  volume    =  64,
  number    =  5,
  pages     = "1644--1650",
  month     =  "15~" # may,
  year      =  1995
}

@ARTICLE{Voronin2001-oc,
  title     = "{Neutron diffraction, synchrotron radiation and EXAFS
               spectroscopy study of crystal structure peculiarities of the
               lanthanum nickelates {La}$_{n+1}${Ni}$_{n}${O}$_{y }$ ($n$=1,2,3)}",
  author    = "Voronin, V I and Berger, I F and Cherepanov, V A and Gavrilova, L
               Ya and Petrov, A N and Ancharov, A I and Tolochko, B P and
               Nikitenko, S G",
  journal   = "Nucl. Instrum. Methods Phys. Res. A",
  publisher = "Elsevier BV",
  volume    =  470,
  number    = "1-2",
  pages     = "202--209",
  month     =  "1~" # sep,
  year      =  2001
}

@ARTICLE{Ling2000-uz,
  title     = "{Neutron diffraction study of {La}$_{3}${Ni}$_{2}${O}$_{7}$:
               Structural relationships among $n=$1, 2, and 3 phases
               {La}$_{n+1}${Ni}$_{n}${O}$_{3n+1}$}",
  author    = "Ling, Christopher D and Argyriou, Dimitri N and Wu, Guoqing and
               Neumeier, J J",
  journal   = "J. Solid State Chem.",
  publisher = "Elsevier BV",
  volume    =  152,
  number    =  2,
  pages     = "517--525",
  month     =  "1~" # jul,
  year      =  2000
}

@ARTICLE{Zhang1994-vm,
  title     = "{Synthesis, structure, and properties of the layered perovskite
               {La}$_{3}${Ni}$_{2}${O}$_{7-\delta}$}",
  author    = "Zhang, Z and Greenblatt, M and Goodenough, J B",
  journal   = "J. Solid State Chem.",
  publisher = "Elsevier BV",
  volume    =  108,
  number    =  2,
  pages     = "402--409",
  month     =  "1~" # feb,
  year      =  1994
}

@ARTICLE{Wang2025-tg,
  title     = "{{Temperature-dependent structural evolution of Ruddlesden-Popper
               bilayer nickelate {La}$_{3}${Ni}$_{2}${O}$_7$}}",
  author    = "Wang, Haozhe and Zhou, Haidong and Xie, Weiwei",
  journal   = "Inorg. Chem.",
  publisher = "American Chemical Society (ACS)",
  volume    =  64,
  number    =  2,
  pages     = "828--834",
  month     =  "20~" # jan,
  year      =  2025
}

@ARTICLE{Wang2024-qi,
  title     = "{{Structure responsible for the superconducting state in
               {La}$_{3}${Ni}$_{2}${O}$_7$ at high-pressure and low-temperature
               conditions}}",
  author    = "Wang, Luhong and Li, Yan and Xie, Sheng-Yi and Liu, Fuyang and
               Sun, Hualei and Huang, Chaoxin and Gao, Yang and Nakagawa,
               Takeshi and Fu, Boyang and Dong, Bo and Cao, Zhenhui and Yu,
               Runze and Kawaguchi, Saori I and Kadobayashi, Hirokazu and Wang,
               Meng and Jin, Changqing and Mao, Ho-Kwang and Liu, Haozhe",
  journal   = "J. Am. Chem. Soc.",
  publisher = "American Chemical Society (ACS)",
  volume    =  146,
  number    =  11,
  pages     = "7506--7514",
  month     =  "20~" # mar,
  year      =  2024
}

@ARTICLE{Zhu2024-rn,
  title     = "{{Superconductivity in pressurized trilayer La$_4$Ni$_3$O$_{10-\delta}$ single crystals}}",
  author    = "Zhu, Yinghao and Peng, Di and Zhang, Enkang and Pan, Bingying and
               Chen, Xu and Chen, Lixing and Ren, Huifen and Liu, Feiyang and
               Hao, Yiqing and Li, Nana and Xing, Zhenfang and Lan, Fujun and
               Han, Jiyuan and Wang, Junjie and Jia, Donghan and Wo, Hongliang
               and Gu, Yiqing and Gu, Yimeng and Ji, Li and Wang, Wenbin and
               Gou, Huiyang and Shen, Yao and Ying, Tianping and Chen, Xiaolong
               and Yang, Wenge and Cao, Huibo and Zheng, Changlin and Zeng,
               Qiaoshi and Guo, Jian-Gang and Zhao, Jun",
  journal   = "Nature",
  publisher = "Springer Science and Business Media LLC",
  volume    =  631,
  number    =  8021,
  pages     = "531--536",
  month     =  "17~" # jul,
  year      =  2024
}

@ARTICLE{Brese1991-mu,
  title     = "{{Bond-valence parameters for solids}}",
  author    = "Brese, N E and O'Keeffe, M",
  journal   = "Acta Crystallogr. B",
  publisher = "International Union of Crystallography (IUCr)",
  volume    =  47,
  number    =  2,
  pages     = "192--197",
  month     =  "1~" # apr,
  year      =  1991
}

@ARTICLE{Huang2026-ax,
  title     = "{{Charge disproportionation driven polar magnetic metallic
               double-layered perovskite {Sr}$_{3}${Co}$_{2}${O}$_7$}}",
  author    = "Huang, Hong-Fei and Sabri, Houssam and Zang, Jiadong and Yu,
               Jie-Xiang",
  journal   = "Chin. Physics Lett.",
  publisher = "IOP Publishing",
  volume    =  43,
  number    =  3,
  pages     =  030712,
  month     =  "1~" # mar,
  year      =  2026
}

@ARTICLE{Zhou2026-uk,
  title     = "{{Geometry-driven polar antiferromagnetic metallicity in a
               double-layered perovskite cobaltate}}",
  author    = "Zhou, Yu and Shu, Xinyu and Zhang, Yang and Liu, Zhiwei and Liu,
               Liangyang and Xiao, Kunhong and Shen, Shengchun and Wu, Sijie and
               Li, Cong and Zhang, Jianbing and Lyu, Yingjie and Wu, Yongshun
               and Sabri, Houssam and Wang, Meng and Yi, Di and Nan, Tianxiang
               and Zhang, Guang-Ming and He, Qing and Zang, Jiadong and Yang,
               Luyi and Zhou, Shuyun and Chen, Hanghui and Yu, Pu",
  journal   = "Nat. Mater.",
  publisher = "Springer Science and Business Media LLC",
  volume    =  25,
  number    =  2,
  pages     = "231--237",
  month     =  feb,
  year      =  2026
}

@ARTICLE{Kim2021-vw,
  title     = "{{Hidden charge order in an iron oxide square-lattice compound}}",
  author    = "Kim, Jung-Hwa and Peets, Darren C and Reehuis, Manfred and Adler,
               Peter and Maljuk, Andrey and Ritschel, Tobias and Allison, Morgan
               C and Geck, Jochen and Mardegan, Jose R L and Bereciartua Perez,
               Pablo J and Francoual, Sonia and Walters, Andrew C and Keller,
               Thomas and Abdala, Paula M and Pattison, Philip and Dosanjh,
               Pinder and Keimer, Bernhard",
  journal   = "Phys. Rev. Lett.",
  publisher = "American Physical Society (APS)",
  volume    =  127,
  number    =  9,
  pages     =  097203,
  month     =  "27~" # aug,
  year      =  2021
}

@ARTICLE{Wang2024-dq,
  title     = "{{Bulk high-temperature superconductivity in pressurized
               tetragonal {La}$_{2}${PrNi}$_{2}${O}$_7$}}",
  author    = "Wang, Ningning and Wang, Gang and Shen, Xiaoling and Hou, Jun and
               Luo, Jun and Ma, Xiaoping and Yang, Huaixin and Shi, Lifen and
               Dou, Jie and Feng, Jie and Yang, Jie and Shi, Yunqing and Ren,
               Zhian and Ma, Hanming and Yang, Pengtao and Liu, Ziyi and Liu,
               Yue and Zhang, Hua and Dong, Xiaoli and Wang, Yuxin and Jiang,
               Kun and Hu, Jiangping and Nagasaki, Shoko and Kitagawa, Kentaro
               and Calder, Stuart and Yan, Jiaqiang and Sun, Jianping and Wang,
               Bosen and Zhou, Rui and Uwatoko, Yoshiya and Cheng, Jinguang",
  journal   = "Nature",
  publisher = "Springer Science and Business Media LLC",
  volume    =  634,
  number    =  8034,
  pages     = "579--584",
  month     =  "2~" # oct,
  year      =  2024
}

@ARTICLE{Li2026-go,
  title     = "{{Bulk superconductivity up to 96 K in pressurized nickelate
               single crystals}}",
  author    = "Li, Feiyu and Xing, Zhenfang and Peng, Di and Dou, Jie and Guo,
               Ning and Ma, Liang and Zhang, Yulin and Wang, Lingzhen and Luo,
               Jun and Yang, Jie and Zhang, Jian and Chang, Tieyan and Chen,
               Yu-Sheng and Cai, Weizhao and Cheng, Jinguang and Wang, Yuzhu and
               Liu, Yuxin and Luo, Tao and Hirao, Naohisa and Matsuoka, Takahiro
               and Kadobayashi, Hirokazu and Zeng, Zhidan and Zheng, Qiang and
               Zhou, Rui and Zeng, Qiaoshi and Tao, Xutang and Zhang, Junjie",
  journal   = "Nature",
  publisher = "Springer Science and Business Media LLC",
  volume    =  649,
  number    =  8098,
  pages     = "871--878",
  month     =  jan,
  year      =  2026
}

@ARTICLE{Lei2018-lg,
  title     = "{{Observation of quasi-two-dimensional polar domains and
               ferroelastic switching in a metal, {Ca$_3$Ru$_2$O$_7$}}}",
  author    = "Lei, Shiming and Gu, Mingqiang and Puggioni, Danilo and Stone,
               Greg and Peng, Jin and Ge, Jianjian and Wang, Yu and Wang,
               Baoming and Yuan, Yakun and Wang, Ke and Mao, Zhiqiang and
               Rondinelli, James M and Gopalan, Venkatraman",
  journal   = "Nano Lett.",
  publisher = "American Chemical Society (ACS)",
  volume    =  18,
  number    =  5,
  pages     = "3088--3095",
  month     =  "9~" # may,
  year      =  2018
}

@ARTICLE{Puphal2024-pw,
  title     = "{{Unconventional crystal structure of the high-pressure
               superconductor {La}$_{3}${Ni}$_{2}${O}$_7$}}",
  author    = "Puphal, P and Reiss, P and Enderlein, N and Wu, Y-M and
               Khaliullin, G and Sundaramurthy, V and Priessnitz, T and Knauft,
               M and Suthar, A and Richter, L and Isobe, M and van Aken, P A and
               Takagi, H and Keimer, B and Suyolcu, Y E and Wehinger, B and
               Hansmann, P and Hepting, M",
  journal   = "Phys. Rev. Lett.",
  publisher = "American Physical Society (APS)",
  volume    =  133,
  number    =  14,
  pages     =  146002,
  month     =  "4~" # oct,
  year      =  2024
}

@ARTICLE{Wang2024-ii,
  title     = "{{Long-range structural order in a hidden phase of
               Ruddlesden-Popper bilayer nickelate {La$_3$Ni$_2$O$_7$}}}",
  author    = "Wang, Haozhe and Chen, Long and Rutherford, Aya and Zhou, Haidong
               and Xie, Weiwei",
  journal   = "Inorg. Chem.",
  publisher = "American Chemical Society (ACS)",
  volume    =  63,
  number    =  11,
  pages     = "5020--5026",
  month     =  "18~" # mar,
  year      =  2024
}

@ARTICLE{Li2024-vg,
  title     = "{{Design and synthesis of three-dimensional hybrid
               Ruddlesden-Popper nickelate single crystals}}",
  author    = "Li, Feiyu and Guo, Ning and Zheng, Qiang and Shen, Yang and Wang,
               Shilei and Cui, Qihui and Liu, Chao and Wang, Shanpeng and Tao,
               Xutang and Zhang, Guang-Ming and Zhang, Junjie",
  journal   = "Phys. Rev. Mater.",
  publisher = "American Physical Society (APS)",
  volume    =  8,
  number    =  5,
  pages     =  053401,
  month     =  "8~" # may,
  year      =  2024
}

@ARTICLE{Chen2024-dr,
  title     = "{{Polymorphism in the Ruddlesden-Popper nickelate La$_3$Ni$_2$O$_7$: Discovery of a hidden phase with distinctive layer stacking}}",
  author    = "Chen, Xinglong and Zhang, Junjie and Thind, Arashdeep S and
               Sharma, Shekhar and LaBollita, Harrison and Peterson, Gordon and
               Zheng, Hong and Phelan, Daniel P and Botana, Antia S and Klie,
               Robert F and Mitchell, J F",
  journal   = "J. Am. Chem. Soc.",
  publisher = "American Chemical Society (ACS)",
  volume    =  146,
  number    =  6,
  pages     = "3640--3645",
  month     =  "14~" # feb,
  year      =  2024
}

@ARTICLE{Moritomo1996-vz,
  title     = "{{Giant magnetoresistance of manganese oxides with a layered
               perovskite structure}}",
  author    = "Moritomo, Y and Asamitsu, A and Kuwahara, H and Tokura, Y",
  journal   = "Nature",
  publisher = "Springer Science and Business Media LLC",
  volume    =  380,
  number    =  6570,
  pages     = "141--144",
  month     =  mar,
  year      =  1996
}

@ARTICLE{Benedek2011-kv,
  title     = "{{Hybrid improper ferroelectricity: a mechanism for controllable
               polarization-magnetization coupling}}",
  author    = "Benedek, Nicole A and Fennie, Craig J",
  journal   = "Phys. Rev. Lett.",
  publisher = "American Physical Society (APS)",
  volume    =  106,
  number    =  10,
  pages     =  107204,
  month     =  "11~" # mar,
  year      =  2011
}

@ARTICLE{Smith2019-nn,
  title     = "{{Infrared nano-spectroscopy of ferroelastic domain walls in
               hybrid improper ferroelectric {Ca}$_{3}${Ti}$_{2}${O}$_7$}}",
  author    = "Smith, K A and Nowadnick, E A and Fan, S and Khatib, O and Lim, S
               J and Gao, B and Harms, N C and Neal, S N and Kirkland, J K and
               Martin, M C and Won, C J and Raschke, M B and Cheong, S-W and
               Fennie, C J and Carr, G L and Bechtel, H A and Musfeldt, J L",
  journal   = "Nat. Commun.",
  publisher = "Springer Science and Business Media LLC",
  volume    =  10,
  number    =  1,
  pages     =  5235,
  month     =  "20~" # nov,
  year      =  2019
}

@ARTICLE{Lee2013-tj,
  title     = "{{Exploiting dimensionality and defect mitigation to create
               tunable microwave dielectrics}}",
  author    = "Lee, Che-Hui and Orloff, Nathan D and Birol, Turan and Zhu, Ye
               and Goian, Veronica and Rocas, Eduard and Haislmaier, Ryan and
               Vlahos, Eftihia and Mundy, Julia A and Kourkoutis, Lena F and
               Nie, Yuefeng and Biegalski, Michael D and Zhang, Jingshu and
               Bernhagen, Margitta and Benedek, Nicole A and Kim, Yongsam and
               Brock, Joel D and Uecker, Reinhard and Xi, X X and Gopalan,
               Venkatraman and Nuzhnyy, Dmitry and Kamba, Stanislav and Muller,
               David A and Takeuchi, Ichiro and Booth, James C and Fennie, Craig
               J and Schlom, Darrell G",
  journal   = "Nature",
  publisher = "Springer Science and Business Media LLC",
  volume    =  502,
  number    =  7472,
  pages     = "532--536",
  month     =  "24~" # oct,
  year      =  2013
}

@ARTICLE{Oh2015-vd,
  title     = "{{Experimental demonstration of hybrid improper ferroelectricity
               and the presence of abundant charged walls in
               (Ca,Sr)$_{3}${Ti}$_{2}${O}$_7$ crystals}}",
  author    = "Oh, Yoon Seok and Luo, Xuan and Huang, Fei-Ting and Wang, Yazhong
               and Cheong, Sang-Wook",
  journal   = "Nat. Mater.",
  publisher = "Springer Science and Business Media LLC",
  volume    =  14,
  number    =  4,
  pages     = "407--413",
  month     =  "12~" # apr,
  year      =  2015
}

@ARTICLE{Liu2018-jx,
  title     = "{{Direct observation of ferroelectricity in
               {Ca}$_{3}${Mn}$_{2}${O}$_7$ and its prominent light absorption}}",
  author    = "Liu, Meifeng and Zhang, Yang and Lin, Ling-Fang and Lin, Lin and
               Yang, Shengwei and Li, Xiang and Wang, Yu and Li, Shaozhen and
               Yan, Zhibo and Wang, Xiuzhang and Li, Xiao-Guang and Dong, Shuai
               and Liu, Jun-Ming",
  journal   = "Appl. Phys. Lett.",
  publisher = "AIP Publishing",
  volume    =  113,
  number    =  2,
  pages     =  022902,
  month     =  "9~" # jul,
  year      =  2018
}

@ARTICLE{Liu2026-vk,
  title         = "{{Density-wave phases, anisotropic transport, and Planckian
                   dissipation in single crystals of the superconductor
                   {La$_3$Ni$_2$O$_7$}}}",
  author        = "Liu, Zhehong and Nakajima, Masamichi and Kriener, Markus and
                   Kitou, Shunsuke and Lyu, Xiaowei and Terakura, Chieko and
                   Karube, Kosuke and Yip, Ka Man and Lazar, Sorin and
                   Nakanishi, Nobuto and Shimada, Keiko and Kikkawa, Akiko and
                   Fujishiro, Yukako and Yu, Xiuzhen and Arima, Taka-Hisa and
                   Tokura, Yoshinori and Taguchi, Yasujiro",
  journal       = "arXiv:2607.26990 [cond-mat.supr-con]",
  year          =  2026,
}

@ARTICLE{Foyevtsova2025-fs,
  title     = "{{Charge distribution and magnetism in bilayer La$_3$Ni$_2$O$_7$: A
               hybrid functional study}}",
  author    = "Foyevtsova, Kateryna and Elfimov, Ilya and Sawatzky, George A",
  journal   = "Phys. Rev. B.",
  publisher = "American Physical Society (APS)",
  volume    =  112,
  number    =  23,
  pages     =  235163,
  month     =  "23~" # dec,
  year      =  2025
}

@ARTICLE{Jiang2025-bu,
  title     = "{{Intertwined charge and spin instability of {La$_3$Ni$_2$O$_7$}}}",
  author    = "Jiang, Guiwen and Qin, Chenye and Foyevtsova, Kateryna and Si,
               Liang and Berciu, Mona and Sawatzky, George A and Jiang, Mi",
  journal   = "Sci. China Phys. Mech. Astron.",
  publisher = "Springer Science and Business Media LLC",
  volume    =  68,
  number    =  9,
  pages     =  297411,
  month     =  "25~" # sep,
  year      =  2025
}

@ARTICLE{Johnston2014-jt,
  title     = "{{Charge disproportionation without charge transfer in the
               rare-earth-element nickelates as a possible mechanism for the
               metal-insulator transition}}",
  author    = "Johnston, Steve and Mukherjee, Anamitra and Elfimov, Ilya and
               Berciu, Mona and Sawatzky, George A",
  journal   = "Phys. Rev. Lett.",
  publisher = "American Physical Society (APS)",
  volume    =  112,
  number    =  10,
  pages     =  106404,
  month     =  "14~" # mar,
  year      =  2014
}

@ARTICLE{Lu2018-fd,
  title     = "{{Site-selective probe of magnetic excitations in rare-earth
               nickelates using resonant inelastic X-ray scattering}}",
  author    = "Lu, Y and Betto, D and Fürsich, K and Suzuki, H and Kim, H-H and
               Cristiani, G and Logvenov, G and Brookes, N B and Benckiser, E
               and Haverkort, M W and Khaliullin, G and Le Tacon, M and Minola,
               M and Keimer, B",
  journal   = "Phys. Rev. X.",
  publisher = "American Physical Society (APS)",
  volume    =  8,
  number    =  3,
  pages     =  031014,
  month     =  "18~" # jul,
  year      =  2018
}

@ARTICLE{Green2016-du,
  title     = "{{Bond disproportionation and dynamical charge fluctuations in the
               perovskite rare-earth nickelates}}",
  author    = "Green, R J and Haverkort, M W and Sawatzky, G A",
  journal   = "Phys. Rev. B.",
  publisher = "American Physical Society (APS)",
  volume    =  94,
  number    =  19,
  pages     =  195127,
  month     =  "15~" # nov,
  year      =  2016
}

@ARTICLE{Chen2024-rixs,
  title     = "{{Electronic and magnetic excitations in {La$_3$Ni$_2$O$_7$}}}",
  author    = "Chen, Xiaoyang and Choi, Jaewon and Jiang, Zhicheng and Mei,
               Jiong and Jiang, Kun and Li, Jie and Agrestini, Stefano and
               Garcia-Fernandez, Mirian and Sun, Hualei and Huang, Xing and
               Shen, Dawei and Wang, Meng and Hu, Jiangping and Lu, Yi and Zhou,
               Ke-Jin and Feng, Donglai",
  journal   = "Nat. Commun.",
  publisher = "Springer Science and Business Media LLC",
  volume    =  15,
  number    =  1,
  pages     =  9597,
  month     =  "6~" # nov,
  year      =  2024
}

@ARTICLE{Takegami2024-fs,
  title     = "{{Absence of Ni$2+$/Ni$3+$ charge disproportionation and possible
               roles of O $2p$ holes in La$_3$Ni$_2$O$_{7-\delta}$ revealed by hard x-ray
               photoemission spectroscopy}}",
  author    = "Takegami, D and Fujinuma, K and Nakamura, R and Yoshimura, M and
               Tsuei, K-D and Wang, G and Wang, N N and Cheng, J-G and Uwatoko,
               Y and Mizokawa, T",
  journal   = "Phys. Rev. B.",
  publisher = "American Physical Society (APS)",
  volume    =  109,
  number    =  12,
  pages     =  125119,
  month     =  "11~" # mar,
  year      =  2024
}

@ARTICLE{Plokhikh2025-main,
  title         = "{{Unraveling Spin Density Wave Order in Layered Nickelates
                   $\mathrm{La_3Ni_2O_7}$ and $\mathrm{La_2PrNi_2O_7}$ via
                   Neutron Diffraction}}",
  author        = "Plokhikh, Igor and Hicken, Thomas J and Keller, Lukas and
                   Pomjakushin, Vladimir and Moody, Samuel H and
                   Foury-Leylekian, Pascale and Krieger, Jonas J and Luetkens,
                   Hubertus and Guguchia, Zurab and Khasanov, Rustem and
                   Gawryluk, Dariusz Jakub",
  journal       = "arXiv:2503.05287 [cond-mat.supr-con]",
  month         =  "7~" # mar,
  year          =  2025,
}

@ARTICLE{Chen2026-main,
  title         = "{{Nature of magnetism in bilayer nickelate La$_3$Ni$_2$O$_7$ single
                   crystals}}",
  author        = "Chen, Lixing and Zhang, Enkang and Hao, Yiqing and Zhu,
                   Yinghao and Cui, Bingkun and Abernathy, Douglas L and
                   Williams, Travis J and Ikeda, Yoichi and Zhang, Hao and Liu,
                   Feiyang and Wang, Wenbin and Wang, Qisi and Zhao, Jun",
  journal       = "arXiv:2605.03448 [cond-mat.str-el]",
  month         =  "5~" # may,
  year          =  2026,
}

@ARTICLE{Khasanov2025-main,
  title     = "{{Pressure-enhanced splitting of density wave transitions in
               {La$_3$Ni$_2$O$_{7–\delta}$}}}",
  author    = "Khasanov, Rustem and Hicken, Thomas J and Gawryluk, Dariusz J and
               Sazgari, Vahid and Plokhikh, Igor and Sorel, Loïc Pierre and
               Bartkowiak, Marek and Bötzel, Steffen and Lechermann, Frank and
               Eremin, Ilya M and Luetkens, Hubertus and Guguchia, Zurab",
  journal   = "Nat. Phys.",
  publisher = "Springer Science and Business Media LLC",
  volume    =  21,
  number    =  3,
  pages     = "430--436",
  month     =  "17~" # mar,
  year      =  2025
}

@ARTICLE{Sasaki1997-qj,
  title     = "{{Structural studies on the phase transition of La$_3$Ni$_2$O$_{6.92}$ at
               about $550\,$K}}",
  author    = "Sasaki, Hisashi and Harashina, Hiroshi and Taniguchi, Satoshi and
               Kasai, Mayumi and Kobayashi, Yoshiaki and Sato, Masatoshi and
               Kobayashi, Takashi and Ikeda, Tomohiro and Takata, Masaki and
               Sakata, Makoto",
  journal   = "J. Phys. Soc. Jpn.",
  publisher = "Physical Society of Japan",
  volume    =  66,
  number    =  6,
  pages     = "1693--1697",
  month     =  "15~" # jun,
  year      =  1997
}

@ARTICLE{Alonso1999-ld,
  title     = "{{Charge disproportionation in $R$NiO$_3$ Perovskites: Simultaneous metal-insulator and structural transition in YNiO$_3$}}",
  author    = "Alonso, J A and García-Muñoz, J L and Fernández-Díaz, M T and
               Aranda, M A G and Martínez-Lope, M J and Casais, M T",
  journal   = "Phys. Rev. Lett.",
  publisher = "American Physical Society (APS)",
  volume    =  82,
  number    =  19,
  pages     = "3871--3874",
  month     =  "10~" # may,
  year      =  1999
}

@ARTICLE{LaBollita2024-rd,
  title     = "{{Assessing spin-density wave formation in La$_3$Ni$_2$O$_7$ from electronic structure calculations}}",
  author    = "LaBollita, Harrison and Pardo, Victor and Norman, Michael R and
               Botana, Antia S",
  journal   = "Phys. Rev. Mater.",
  publisher = "American Physical Society (APS)",
  volume    =  8,
  number    =  11,
  pages     = "L111801",
  month     =  "20~" # nov,
  year      =  2024
}

@ARTICLE{Zhou2025-main,
  title     = "{{Investigations of key issues on the reproducibility of
               high-$T_c$ superconductivity emerging from compressed
               {La}$_{3}${Ni}$_{2}${O}$_{7}$}}",
  author    = "Zhou, Yazhou and Guo, Jing and Cai, Shu and Sun, Hualei and Li,
               Chengyu and Zhao, Jinyu and Wang, Pengyu and Han, Jinyu and Chen,
               Xintian and Chen, Yongjin and Wu, Qi and Ding, Yang and Xiang,
               Tao and Mao, Ho-Kwang and Sun, Liling",
  journal   = "Matter Radiat. Extrem.",
  publisher = "AIP Publishing",
  volume    =  10,
  number    =  2,
  pages     =  027801,
  month     =  "1~" # mar,
  year      =  2025
}

@ARTICLE{Yashima2026-qv,
  title     = "{{Large-tiny spin-stripe SDW order in La$_{3}$Ni$_{2}$O$_{7}$
               via $^{139}$ La-NQR spectroscopy: Reconsideration based on the
               \textit{Am}2\textit{m} structure}}",
  author    = "Yashima, Mitsuharu and Lee, Hongnam and Ino, Takumi and Seto,
               Nina and Kakoi, Masataka and Sakurai, Hiroya and Takano,
               Yoshihiko and Mukuda, Hidekazu",
  journal   = "J. Phys. Soc. Jpn.",
  publisher = "Physical Society of Japan",
  volume    =  95,
  number    =  12,
  month     =  "15~" # dec,
  year      =  2026
}

@ARTICLE{Dong2025-ni,
  title     = "{{Interstitial oxygen order and its competition with
               superconductivity in {La}$_{2}${PrNi}$_{2}${O}$_{7+\delta}$}}",
  author    = "Dong, Zehao and Wang, Gang and Wang, Ningning and Dong, Wen-Han
               and Gu, Lin and Xu, Yong and Cheng, Jinguang and Chen, Zhen and
               Wang, Yayu",
  journal   = "Nat. Mater.",
  publisher = "Springer Science and Business Media LLC",
  volume    =  24,
  number    =  12,
  pages     = "1927--1934",
  month     =  "15~" # dec,
  year      =  2025
}

@ARTICLE{Ivlev2019-el,
  title     = "{{HKLF5Tools: a program for processing diffraction data of
               non-merohedrally twinned crystals}}",
  author    = "Ivlev, Sergei I and Conrad, Matthias and Kraus, Florian",
  journal   = "Z. Kristallogr. Cryst. Mater.",
  publisher = "Walter de Gruyter GmbH",
  volume    =  234,
  number    =  6,
  pages     = "415--418",
  month     =  "27~" # may,
  year      =  2019
}

@ARTICLE{Shi2025-qr,
  title     = "{{Spin density wave rather than tetragonal structure is
               prerequisite for superconductivity in La$_3$Ni$_2$O$_{7-\delta}$}}",
  author    = "Shi, Mengzhu and Peng, Di and Li, Yikang and Yang, Shaohua and
               Xing, Zhenfang and Wang, Yuzhu and Fan, Kaibao and Li, Houpu and
               Wu, Rongqi and Ge, Binghui and Zeng, Zhidan and Zeng, Qiaoshi and
               Ying, Jianjun and Wu, Tao and Chen, Xianhui",
  journal   = "Nat. Commun.",
  publisher = "Springer Science and Business Media LLC",
  volume    =  16,
  number    =  1,
  pages     =  9141,
  month     =  "15~" # oct,
  year      =  2025
}

\newpage

%%===========================================================================================%%
%% MAIN TEXT FIGURES
%%===========================================================================================%%
\clearpage

\begin{center}
\Large{Main Text Figures}
\end{center}
\FloatBarrier
\vspace{5mm}
\clearpage

\clearpage

\clearpage
\newpage
\begin{table}[ht]
\centering
\caption{\textbf{Refinement statistics for \textit{Amam} and \textit{Am}2\textit{m} models of \LNO{}.} Oxygen occupancies are fixed to $1$, since the refinement results in the stoichiometric values for the $Am2m$ structure; see Supplementary Note~\ref{sec:SI_str}.}
\label{Tab:refinement}
\setlength{\tabcolsep}{6pt}
\begin{tabular}{c c c}
\hline
Space group & $Amam$ (No.~63) & $Am2m$ (No.~38)\\
\hline
Temperature (K) & \multicolumn{2}{c}{$50$} \\
Wavelength ($\text{\AA}$) & \multicolumn{2}{c}{$0.343008$} \\
$a$ ($\text{\AA}$) & \multicolumn{2}{c}{$5.37640(10)$} \\
$b$ ($\text{\AA}$) & \multicolumn{2}{c}{$5.45650(10)$} \\
$c$ ($\text{\AA}$) & \multicolumn{2}{c}{$20.5013(4)$} \\
$\alpha = \beta = \gamma$ ($^\circ$) & \multicolumn{2}{c}{$90$} \\
Cell volume ($\text{\AA}^{3}$) & \multicolumn{2}{c}{$601.43(2)$} \\
$Z$ & \multicolumn{2}{c}{$4$} \\
$F(000)$ & \multicolumn{2}{c}{$1132$} \\
$d_\text{min}$ (\AA) & \multicolumn{2}{c}{$0.3249$} \\
Completeness (\%) & $98$ & $97$ \\
Redundancy & $16.296$ & $8.527$ \\
$N_\mathrm{total}$ & \multicolumn{2}{c}{$92383$} \\
$N_\mathrm{unique}$ & $5542$ & $10834$ \\
\hline
Inversion twin fractions & Not applicable & $0.63(2)$, $0.37(2)$ \\
$N_\mathrm{parameters}$ & $37$ & $72$ \\
$R_1\ (I>3\sigma(I)\ /\ \text{all})$ & $4.19\ \%$ / $4.22\ \%$ & $1.24\ \%$ / $1.28\ \%$ \\
$wR(F)\,$ $(I>3\sigma(I)\ /\ \text{all})$ & $10.39\ \%$ / $10.40\ \%$ & $1.73\ \%$ / $1.75\ \%$ \\
GOF $(I>3\sigma(I)\ /\ \text{all})$ & $7.20$ / $7.13$ & $1.07$ / $1.07$ \\
\hline
\end{tabular}
\end{table}

\clearpage

\clearpage
\newpage

%%%%%%%%%%%%%%%%%%%%%%%%%%%%%%%%%%%%%%%%%%%%%%%%%%%%%%%%%%%%%%%%%%%%%%%
%%%%%%%%%%%%%%%%%%%%%%%%%%%%%%%%%%%%%%%%%%%%%%%%%%%%%%%%%%%%%%%%%%%%%%%
%%				METHODS
%%%%%%%%%%%%%%%%%%%%%%%%%%%%%%%%%%%%%%%%%%%%%%%%%%%%%%%%%%%%%%%%%%%%%%%
%%%%%%%%%%%%%%%%%%%%%%%%%%%%%%%%%%%%%%%%%%%%%%%%%%%%%%%%%%%%%%%%%%%%%%%
\begin{center}
\Large{Methods}
\end{center}

\textbf{Single crystal growth}\\
Single crystals of \LNO{} were grown by a molten salt flux evaporation method at ambient pressure~\cite{Li2026-go}. High-purity La$_2$O$_3$ ($99.99~\%$) and NiO were used as starting materials. La$_2$O$_3$ was dried at $1000\,^\circ$C overnight, and the oxide powders were weighed and ground. The anhydrous K$_2$CO$_3$ was used as flux and then mixed with the oxide powders in a flux-to-solute mass ratio of $15:1$. To prevent moisture absorption, all the processes were performed inside a glovebox. The mixture was placed in an alumina crucible that was covered with a lid to control the evaporation rate. The crystals were grown in the furnace at a temperature of $1000\text{-}1050\,^\circ$C for $72\,$h while evaporating the flux gradually, and then were cooled to room temperature naturally. The crystals were extracted by soaking the mixture in deionized water. Single crystals of \LPNO{} were grown using a molten-salt flux method similar to that described in Ref.~\cite{Shi2025-qr}. First, lanthanum nitrate hexahydrate, praseodymium(III) nitrate hexahydrate, nickel(II) nitrate hexahydrate, and citric acid (CA) were dissolved in water at a molar ratio of CA:La:Pr:Ni = 7.5:2:1:2. The resulting solution was preheated in an oven at $200\,^\circ$C for approximately $12\,$h. The dried product was subsequently transferred to a muffle furnace, heated to $800\,^\circ$C over $200$ min, and maintained at this temperature for $6\,$h to obtain the precursor (P). The as-prepared precursor was then mixed with a NaCl/KCl flux at a P: NaCl: KCl mass ratio of $1:14:16$ and loaded into a corundum crucible. The crucible was heated in air to $1150\,^\circ$C over $10\,$h and held at this temperature for $48~$h, followed by slow cooling to $1110\,^\circ$C over $7\,$days. Finally, the resulting microcrystals were recovered by washing away the residual salt flux with water.

\bigskip

\textbf{Synchrotron X-ray diffraction}\\
Synchrotron XRD measurements were performed at the bending-magnet beamline BL$02$B$1$ and the undulator beamline BL$05$XU of the SPring-$8$ synchrotron radiation facility (Japan). The X-ray energy was tuned to $36\,$keV and $20\,$keV in these beamlines, respectively. Single crystals
with a size of $\sim$ $20\,\mu$m were employed in the measurements. N$_2$-gas-blowing devices were used for temperatures between $100\,$K and $650\,$K, and a He-gas-blowing device for temperatures below $100\,$K. Bragg reflections were recorded using a CdTe PILATUS area detector in BL$02$B$1$ and a Si EIGERX area detector in BL$05$XU, together with the \textsc{CrysAlisPro} software~\cite{Agilent-Technologies-Ltd2014-hl}. For structure refinement using BL$02$B$1$ data, the reciprocal space was scanned in steps of $\Delta \omega = 0.01^\circ$ for \LNO{} and $\Delta \omega = 0.1^\circ$ for \LPNO{}. Orientation-dependent absorption is taken into account by an automated correction procedure implemented in CrysAlisPro using multiple reflections. On the other hand, we do not apply an additional spherical absorption correction; due to the high X-ray energy of $36\,$keV, the linear attenuation coefficient is relatively small, $\mu=3.677\,\mathrm{mm}^{-1}$, and, with the small single crystal size $\sim 20\,\mu\mathrm{m}$, only about $7\,\%$ of the X-rays are absorbed. This small, nearly uniform absorption can be incorporated into the overall scale factor during refinement. Symmetry-equivalent reflections were averaged, and structure refinements were performed with \textsc{Jana2006}~\cite{Petricek2014-pc}.

In Fig.~\ref{main:Fig1}\textbf{f}, we observe minor diffuse streaks along $c^*$, whose intensity is about $1,600$ times weaker than a Bragg reflection. However, the Bragg reflections have a sharp peak structure, consistent with the high quality of our single crystals. This enables us to perform high-quality structure refinements even when such minor diffuse streaks are present. The streaks thus indicate stacking faults, as common in \LNO{}~\cite{Wang2024-dq}. Indeed, previous scanning transmission electron microscopy measurements reveal similar diffuse streaks due to stacking faults~\cite{Zhou2025-main}.
\bigskip

\textbf{Comparison of structure models in \LNO{}}\\
The highest-symmetry structure of the well-known bilayer Ruddlesden-Popper transition metal oxides corresponds to a monotonic stacking of bilayers (sequence: $2222$) with tetragonal space group $I4/mmm$. As a subgroup of $I4/mmm$, most studies of \LNO{} focused on the $Amam$ structure at ambient pressure~\cite{Zhang1994-vm,Taniguchi1995-lg,Ling2000-uz,Voronin2001-oc,Wang2024-qi,Wang2025-tg}, which is corrected to $Am2m$ in the present work. As discussed in the Main Text, $Am2m$ allows for two independent nickel sites due to charge ordering -- a major difference to $Amam$. On the other hand, polymorphism is also reported in some seminal works on \LNO{} crystals, depending on crystal growth conditions, for example, when using the floating-zone method. The polymorphs with space groups of $Cmmm$~\cite{Chen2024-dr,Wang2024-ii} and $Fmmm$~\cite{Puphal2024-pw} are characterized by a stacking of monolayers and trilayers ($1313$), instead of $2222$. The $Fmmm$ polymorph is known to exhibit stacking disorder, as evidenced by pronounced streaks in synchrotron XRD measurements~\cite{Puphal2024-pw}. Another stacking pattern of $1212$ is also reported~\cite{Li2024-vg}. 
In the present study, we focus on high-quality single crystals with a uniform bilayer stacking sequence, consistent with the best crystals reported in the literature~\cite{Zhang1994-vm,Taniguchi1995-lg,Ling2000-uz,Voronin2001-oc,Wang2024-qi,Wang2025-tg}.
\newpage

\clearpage
\textbf{Acknowledgments}\\
We thank R.~Arita, S.~Suetsugu, M.~Kato, Max~T.~Birch, and M.~Kriener for discussions. This work was supported by JSPS KAKENHI Grants No. JP23H05431, No. JP24H01607, JP24K17006, 24H01644, JP26H00644, JP26H00626, JP26H01288, JP26K00649 and JP26H01290 as well as JST ACT-X Grant No.~JPMJAX26EH and JST CREST Grant No. JPMJCR20T1 (Japan) and JST FOREST Grant No. JPMJFR2238, and No. JPMJFR2362 (Japan). It was also supported by JST as part of Adopting Sustainable Partnerships for Innovative Research Ecosystem (ASPIRE), Grant Number JPMJAP2426. M.H. is supported by the Deutsche Forschungsgemeinschaft (DFG, German Research Foundation) via Transregio TRR 360 – 492547816. J.P.S., N.N.W. and J.G.C. are supported by the National Key R\&D Program of China (2023YFA1406100) and the National Natural Science Foundation of China (12025408, 12404179, 12625401, and 12522407). The synchrotron single-crystal X-ray experiments were performed at BL02B1 and BL05XU of SPring-8 with the approval of RIKEN (Proposal Nos. 2025B1928, 2025B1855, 2025B2178, 2025B2056, and 2026B1655).

\bigskip
\textbf{Author contributions}\\
M.H. conceived the project. J.P.S., Y.Y., L.Z, N.W., and J.G.C. grew and characterized the single crystals. R.M., C.K., S.K., Y.N., K.I, and T.S. performed synchrotron XRD measurements. R.M. analyzed the XRD data in consultation with S.K and T.-h.A. R.M. wrote the manuscript in collaboration with M.H., S.K., and T.-h.A.; all authors discussed the results and commented on the manuscript.\\

\textbf{Competing interests}\\
The authors declare no competing interests.\\

%%===========================================================================================%%
%% Supporting Information
%%===========================================================================================%%

\clearpage
\begin{center}
\Large{Supplementary Note}
\end{center}

\stepcounter{hoge}
\renewcommand\thefigure{S\arabic{figure}} 
\renewcommand\thetable{S\arabic{table}} 
\renewcommand\thesection{S\arabic{section}} 
\renewcommand\theequation{S\arabic{equation}} 

\makeatletter
\renewcommand\@bibitem[1]{\item\if@filesw \immediate\write\@auxout
    {\string\bibcite{#1}{A\the\value{\@listctr}}}\fi\ignorespaces}
\def\@biblabel#1{S[#1]}
\makeatother

%======================%%
% SI SECTION 
%====================%%
\section{\NoCaseChange{Details of synchrotron X-ray diffraction and structure refinement on \LNO{}}}
\label{sec:SI_str}

The diffraction pattern in Fig.~\ref{main:Fig1}\textbf{f} shows diffuse streaks along $c^*$, consistent with stacking faults, likely resulting from intergrowths of different Ruddlesden--Popper (RP) phases~\cite{Wang2024-dq}. To quantify the intensity of the streaks, we illustrate a line cut along $0\,0\,l$ in Supplementary Fig.~\ref{SI:streak}. The streak intensity around $0\,0\,13$ is approximately $1,600$ times weaker than the Bragg reflection $0\,0\,12$, demonstrating the weakness of the diffuse scattering. This supports the minor impact of the intergrowth. On the other hand, the Bragg reflections show a sharp peak structure, in line with the high quality of our single crystals.

Supplementary Fig.~\ref{SI:domain} illustrates the temperature dependence of peak splitting from orthorhombic twins. The two peaks get closer as temperature increases towards $T_\mathrm{CO}=560\,$K. This reflects the temperature dependence of lattice constants $a$ and $b$; their values become closer, but a residual difference remains even above $T_\mathrm{CO}$.

Regarding the oxygen occupancies, their refinement for $Amam$ produces a large deviation from $1$ for O$3$: $1.193(18)$. This results from an attempt to compensate for an unphysical thermal ellipsoid of this structure (Supplementary Fig.~\ref{SI:thermal}\textbf{a},\textbf{b}), supporting our conclusion that the $Amam$ model is incorrect. On the other hand, the refinement of the $Am2m$ model produces sufficiently small, regularly shaped thermal ellipsoids at all the sites, as shown in Supplementary Fig.~\ref{SI:thermal}\textbf{c}. Furthermore, refinement of the oxygen occupancies yields stoichiometric values, attesting to the high quality of our single crystals. Details of the refinement results are provided in Table~\ref{Tab:refinement} and Supplementary Tables~\ref{SI:ref_pos},~\ref{SI:ref_thermal}. In the Main Text, we use the results obtained from refinements with the oxygen occupancies fixed to one.

To characterize charge order, we calculate bond-valence sums (BVS). Empirically, the BVS-derived valence of atom $i$ is given by a sum of the bond valence $v_{ij}$ between $i$ and ligand $j$: 
\begin{equation}
    V_i = \sum_jv_{ij} = \sum_j\exp\left(\frac{R_{ij} - d_{ij}}{b}\right),
\end{equation}
where $R_{ij}$ is the material-dependent bond-valence parameter, $d_{ij}$ is the bond length, and $b$ is the universal constant $0.37~\text{\AA}$. We use $R_{ij}=1.654~\text{\AA}$ for Ni$^{2+}$ and O~\cite{Brese1991-mu}.

\begin{figure}[ht]%
\centering
\includegraphics[width=0.6\textwidth]{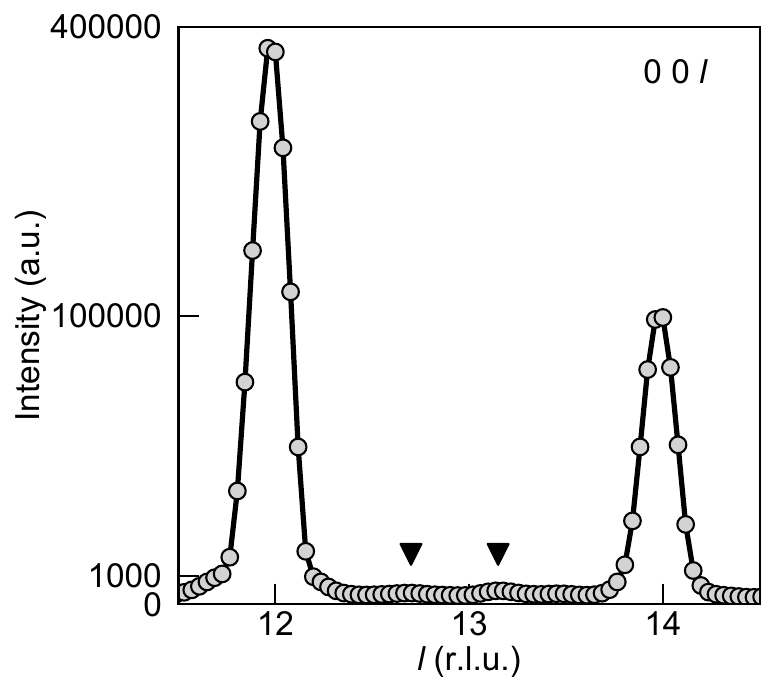}
\caption{\textbf{Minor role of potential stacking faults.} Line cut along $0\,0\,l$ for \LNO{} (piece P1), where diffuse streaks appear in the synchrotron X-ray diffraction data (Fig.~\ref{main:Fig1}\textbf{f}). The integrated intensity for the streak at $0\,0\,13$ is around $1,600$ times weaker than the Bragg reflection $0\,0\,12$. The integrated intensity is presented on a square root scale to emphasize the regime of weak intensities.
}
\label{SI:streak}
\end{figure}

\begin{figure}[ht]%
\centering
\includegraphics[width=1.0\textwidth]{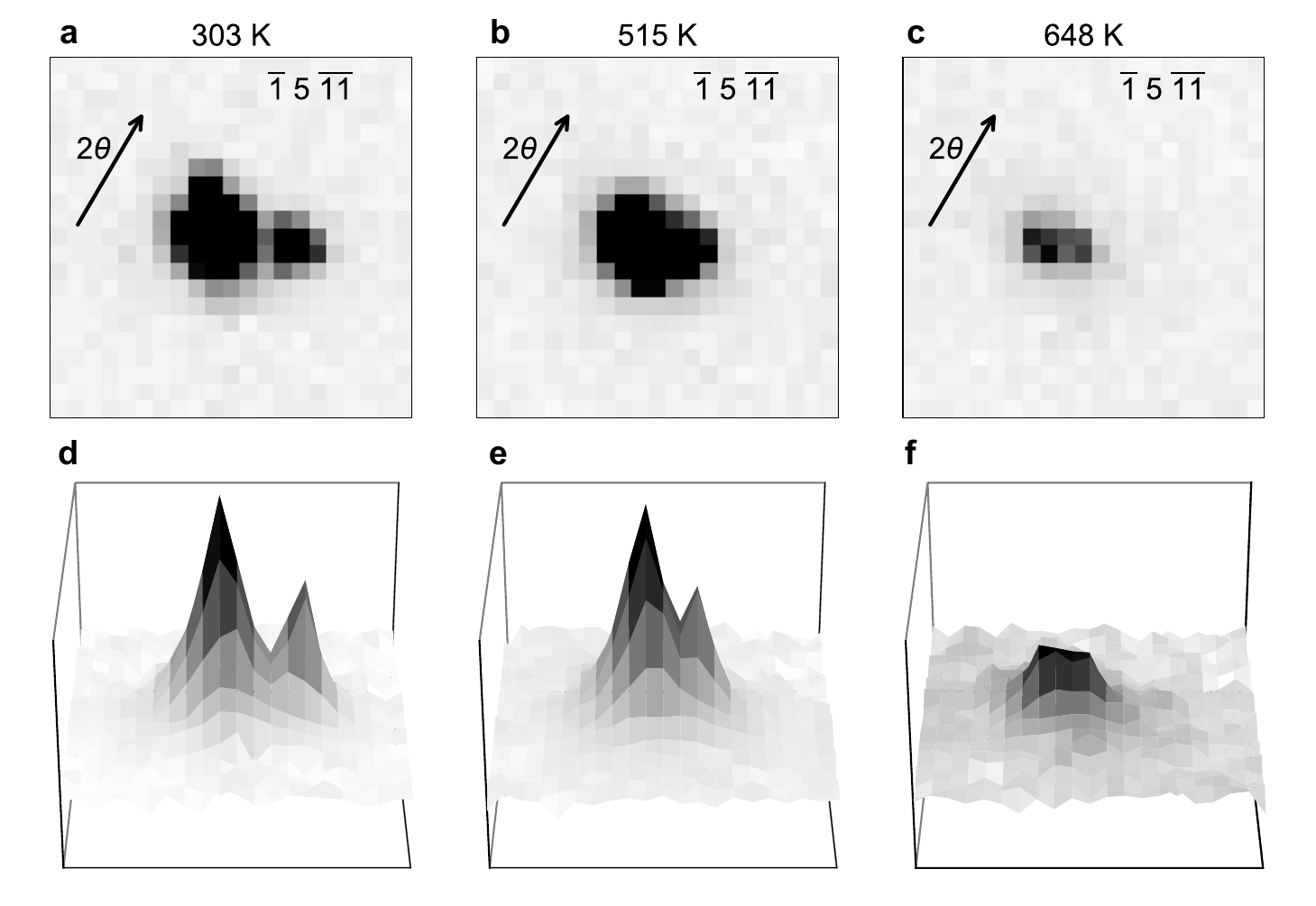}
\caption{\textbf{Temperature dependence of peak splitting associated with orthorhombic twins.}
\textbf{a}-\textbf{c}, X-ray diffraction map around the Bragg reflection $\overline{1}\, 5\, \overline{11}$ in \LNO{} (piece P2) at $303\,$K, $515\,$K, and $648\,$K, respectively. The $d$-spacing of the reflection is $d=0.93\,$\AA. This reflection, corresponding to $2\, 3\, \overline{11}$ in a tetragonal cell, is also allowed in the parent tetragonal lattice. Only two split peaks, corresponding to orthorhombic twins, are resolved. They gradually move closer to each other as temperature increases towards a charge-ordering (CO) transition at $T_\mathrm{CO}=560\,$K. Weak peak splitting still remains above $T_\mathrm{CO}$, as evidenced by a residual difference in the lattice constants $a$ and $b$. \textbf{d}-\textbf{f}, Three-dimensional view of panels~\textbf{a}-\textbf{c}, respectively. Intensities are shown on a logarithmic scale for panels~\textbf{d}-\textbf{f}.
}
\label{SI:domain}
\end{figure}
\FloatBarrier

\begin{figure}[ht]
\centering
\includegraphics[width=0.9\textwidth]{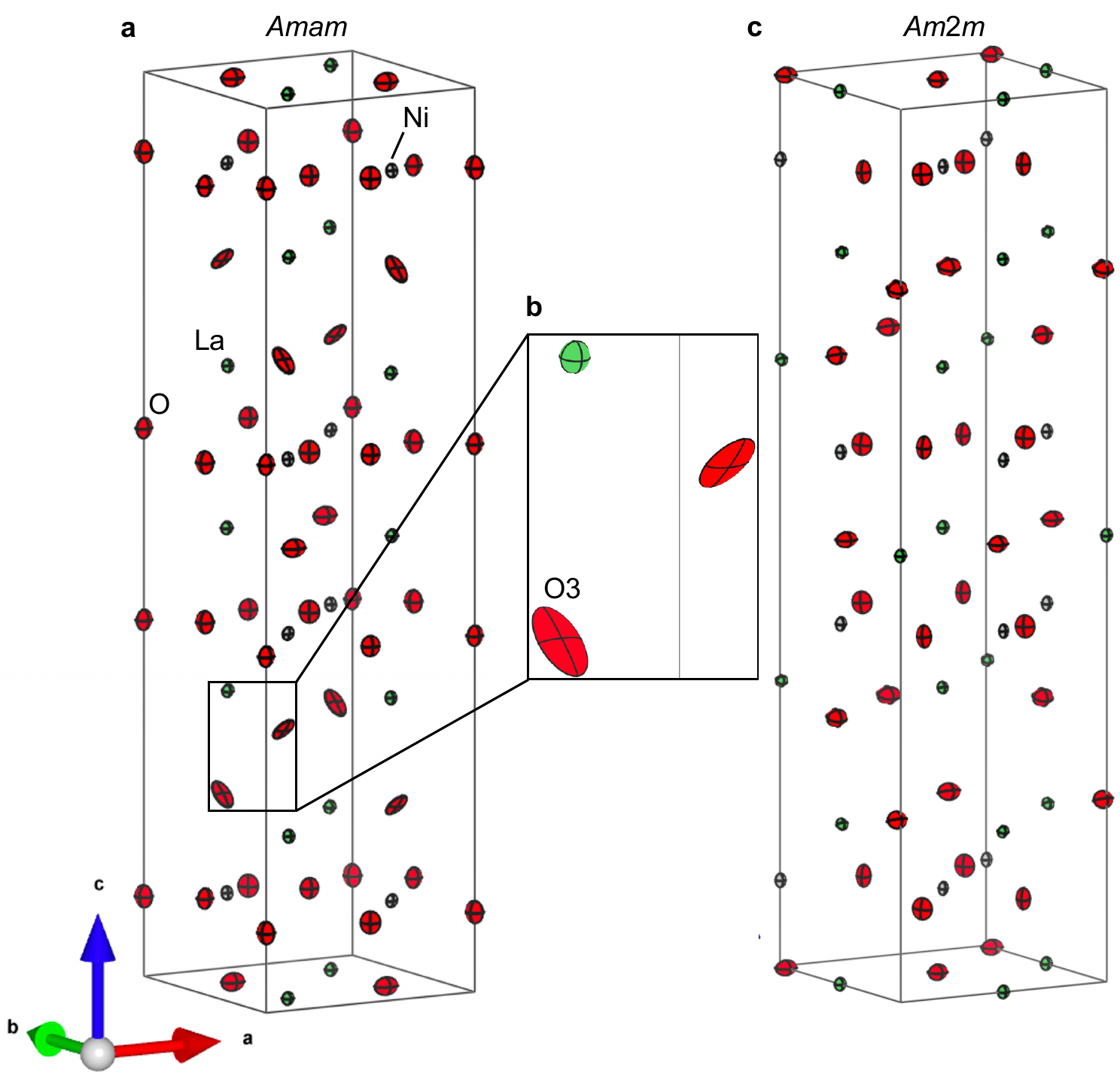}
\caption{\textbf{Anomalous thermal ellipsoids in the \textit{Amam} structure at 50~K.} \textbf{a}, Thermal ellipsoids in the refined $Amam$ structure, drawn at the $99~\%$ probability level. \textbf{b}, Close-up view of panel~\textbf{a}. O$3$ exhibits a flat ellipsoid, indicative of the improper structure model. \textbf{c}, Thermal ellipsoids in the refined $Am2m$ structure. No anomalies are observed at any site. 
}
\label{SI:thermal}
\end{figure}
\FloatBarrier

\begin{table}[htbp]
\centering
\caption{\textbf{Refinement statistics for the charge-disordered phase of \LNO{} ($Amam$) and for the charge-ordered phase of \LPNO{} ($Am2m$).}
Oxygen occupancies are fixed to $1$. For \LPNO{}, structure refinement is performed using high-angle reflections ($\sin\theta/\lambda\geq0.6$), and the statistics are reported for all reflections after the removal of $111$ overlapping low-angle reflections. The redundancy $2.3$ for \LPNO{} is the pre-merging multiplicity reported by CrysAlisPro for the twin dataset exported in HKLF~5 format. $N_{\mathrm{refine}}$ denotes the number of reflections used in the refinement; for \LNO{}, the HKLF~5 reflection file is not used, and $N_{\mathrm{refine}}$ equals the number of independent reflections.}
\label{Tab:SI_refinement}
\setlength{\tabcolsep}{6pt}
\begin{tabular}{c c c}
\hline
 & \LNO{}
 & \LPNO{} \\
\hline
Space group & $Amam$ (No.~63) & $Am2m$ (No.~38) \\
\hline
Temperature (K) & $670$ & $300$ \\
Wavelength ($\text{\AA}$) & $0.32494$ & $0.34354$ \\
$a$ ($\text{\AA}$) & $5.4575(2)$ & $5.37100(10)$ \\
$b$ ($\text{\AA}$) & $5.46990(10)$ & $5.45710(10)$ \\
$c$ ($\text{\AA}$) & $20.5852(5)$ & $20.4353(3)$ \\
$\alpha = \beta = \gamma$ ($^\circ$) & $90$ & $90$ \\
Cell volume ($\text{\AA}^{3}$) & $614.51(3)$ & $598.960(18)$ \\
$Z$ & $4$ & $4$ \\
$F(000)$ & $1132$ & $1139$ \\
$d_{\text{min}}$ (\AA) & $0.2950$ & $0.3101$ \\
Completeness (\%) & $96$ & $95$ \\
Redundancy & $6.306$ & $2.3$ \\
$N_{\mathrm{total}}$  & $40027$ & $17946$ \\
$N_{\mathrm{refine}}$ & $6347$  & $17039$ \\
\hline
Orthorhombic twin fractions
& \(0.577(7)\), \(0.423(7)\) & See Supplementary Table~\ref{Tab:LPNO_twin} \\
$N_{\mathrm{parameters}}$ & $38$ & $76$ \\
$R_1\ (I>3\sigma(I)\ /\ \text{all})$
& $3.98\ \%$ / $7.04\ \%$
& $2.30\ \%$ / $4.42\ \%$ \\
$wR(F)$\ $(I>3\sigma(I)\ /\ \text{all})$
& $5.48\ \%$ / $6.81\ \%$
& $2.80\ \%$ / $2.83\ \%$ \\
GOF $(I>3\sigma(I)\ /\ \text{all})$
& $1.75$ / $1.68$
& $1.21$ / $0.75$ \\
\hline
\end{tabular}
\end{table}
\FloatBarrier

\setlength{\LTcapwidth}{\textwidth}
\begin{longtable}{c c c c c c c}
\caption{\textbf{Atomic fractional coordinates and equivalent isotropic displacement parameters $U_{\mathrm{eq}}$ for \LNO{} and \LPNO{}.} Refinement of oxygen occupancies shows stoichiometric values, consistent with sufficiently small displacement parameters.}
\label{SI:ref_pos}
\label{} \\

\hline
Atom & $x$ & $y$ & $z$ & $U_{\mathrm{eq}}$ (\AA$^2$) & Wyckoff Letters & Occ. \\
\hline
\endfirsthead

\multicolumn{7}{c}%
{{\tablename\ \thetable{} -- continued from previous page}} \\

\hline
Atom & $x$ & $y$ & $z$ & $U_{\mathrm{eq}}$ (\AA$^2$) & Wyckoff Letters & Occ. \\
\hline
\endhead

\hline
\multicolumn{7}{r}{{Continued on next page}} \\
\endfoot

\hline
\endlastfoot

\multicolumn{7}{c}{\LNO{}, $Am$2$m$ ($50\,$K);} oxygen occupancies fixed to $1$ \\
\hline

La1 & $0$          & $-0.012047(7)$  & $0.320215(2)$  & $0.001340(6)$  & $4d$ & $1$ \\
La2 & $0.5$        & $0.006720(7)$   & $0.820477(7)$  & $0.001261(6)$  & $4e$ & $1$ \\
La3 & $0$          & $0.001873(8)$   & $0.5$          & $0.001493(7)$  & $2a$ & $1$ \\
La4 & $0.5$        & $0.003933(8)$   & $0$            & $0.001537(7)$  & $2b$ & $1$ \\
Ni1 & $0$          & $-0.003971(15)$ & $0.096251(4)$  & $0.001462(11)$ & $4d$ & $1$ \\
Ni2 & $0.5$        & $0.001624(16)$  & $0.595733(4)$  & $0.001232(10)$ & $4e$ & $1$ \\
O1  & $0.25487(9)$ & $-0.74434(9)$   & $0.911440(11)$ & $0.00371(3)$   & $8f$ & $1$ \\
O2  & $0.74507(8)$ & $-0.25825(8)$   & $0.894116(12)$ & $0.00377(3)$   & $8f$ & $1$ \\
O3  & $0$          & $0.03234(11)$   & $0.797170(19)$ & $0.00402(6)$   & $4d$ & $1$ \\
O4  & $0.5$        & $-0.04423(11)$  & $0.29401(2)$   & $0.00461(7)$   & $4e$ & $1$ \\
O5  & $0$          & $-0.04841(15)$  & $0$            & $0.00400(9)$   & $2a$ & $1$ \\
O6  & $0.5$        & $0.04793(15)$   & $0.5$          & $0.00364(8)$   & $2b$ & $1$ \\

\hline
\multicolumn{7}{c}{\LNO{}, $Am$2$m$ ($50\,$K);} oxygen occupancies refined \\
\hline

La1 & $0$          & $-0.012049(7)$  & $0.320216(2)$  & $0.001337(6)$  & $4d$ & $1$          \\
La2 & $0.5$        & $0.006723(7)$   & $0.820476(7)$  & $0.001258(6)$  & $4e$ & $1$          \\
La3 & $0$          & $0.001874(8)$   & $0.5$          & $0.001490(7)$  & $2a$ & $1$          \\
La4 & $0.5$        & $0.003931(8)$   & $0$            & $0.001534(7)$  & $2b$ & $1$          \\
Ni1 & $0$          & $-0.003970(16)$ & $0.096253(4)$  & $0.001460(11)$ & $4d$ & $1$          \\
Ni2 & $0.5$        & $0.001626(16)$  & $0.595733(4)$  & $0.001229(10)$ & $4e$ & $1$          \\
O1  & $0.25485(9)$ & $-0.74435(9)$   & $0.911429(12)$ & $0.00376(3)$   & $8f$ & $1.0057(19)$ \\
O2  & $0.74507(8)$ & $-0.25825(8)$   & $0.894128(13)$ & $0.00386(3)$   & $8f$ & $1.0087(18)$ \\
O3  & $0$          & $0.03231(11)$   & $0.79718(2)$   & $0.00415(7)$   & $4d$ & $1.010(3)$   \\
O4  & $0.5$        & $-0.04420(12)$  & $0.29401(2)$   & $0.00451(7)$   & $4e$ & $0.995(3)$   \\
O5  & $0$          & $-0.04838(15)$  & $0$            & $0.00413(9)$   & $2a$ & $1.011(3)$   \\
O6  & $0.5$        & $0.04786(15)$   & $0.5$          & $0.00371(9)$   & $2b$ & $1.007(3)$   \\

\hline
\multicolumn{7}{c}{\LNO{}, $Amam$ ($670\,$K)} \\
\hline

La1 & $0.75$ & $0.24310(5)$ & $0.820549(4)$  & $0.01159(3)$ & $8g$ & $1$ \\
La2 & $0.25$ & $0.74944(2)$ & $1$            & $0.01528(6)$ & $4c$ & $1$ \\
Ni1 & $0.25$ & $0.25181(3)$ & $0.903669(9)$  & $0.00824(4)$ & $8g$ & $1$ \\
O1  & $0.75$ & $0.2783(3)$  & $0.70423(8)$   & $0.0225(4)$  & $8g$ & $1$ \\
O2  & $0.5$  & $0$          & $0.90905(7)$   & $0.0180(3)$  & $8e$ & $1$ \\
O3  & $0.5$  & $0.5$        & $0.89660(10)$  & $0.0189(3)$  & $8e$ & $1$ \\
O4  & $0.25$ & $0.2868(5)$  & $1$            & $0.0205(5)$  & $4c$ & $1$ \\

\hline
\multicolumn{7}{c}{La$_2$PrNi$_2$O$_7$, $Am2m$($300\,$K)} \\
\hline

La1 & $0$            & $-0.00923(6)$   & $0.31983(2)$ & $0.00495(7)$  & $4d$ & $0.64(3)$ \\
Pr1 & $0$            & $-0.00923(6)$   & $0.31983(2)$ & $0.00495(7)$  & $4d$ & $0.36(3)$ \\
La2 & $0.5$          & $0.01180(5)$    & $0.81984(2)$ & $0.00567(8)$  & $4e$ & $0.64(3)$ \\
Pr2 & $0.5$          & $0.01180(5)$    & $0.81984(2)$ & $0.00567(8)$  & $4e$ & $0.36(3)$ \\
La3 & $0$            & $0.00312(6)$    & $0.5$        & $0.00681(10)$ & $2a$ & $0.89(3)$ \\
Pr3 & $0$            & $0.00312(6)$    & $0.5$        & $0.00681(10)$ & $2a$ & $0.11(3)$ \\
La4 & $0.5$          & $0.00638(7)$    & $0$          & $0.00653(10)$ & $2b$ & $0.89(3)$ \\
Pr4 & $0.5$          & $0.00638(7)$    & $0$          & $0.00653(10)$ & $2b$ & $0.11(3)$ \\
Ni1 & $0$            & $-0.00106(19)$  & $0.09649(5)$ & $0.00388(14)$ & $4d$ & $1$       \\
Ni2 & $0.5$          & $0.00530(19)$   & $0.59646(5)$ & $0.00383(14)$ & $4e$ & $1$       \\
O1  & $0.2534(13)$   & $-0.7498(10)$   & $0.91126(7)$ & $0.0098(2)$   & $8f$ & $1$       \\
O2  & $0.7481(16)$   & $-0.2515(10)$   & $0.89218(8)$ & $0.0103(2)$   & $8f$ & $1$       \\
O3  & $0$            & $0.0466(9)$     & $0.7952(2)$  & $0.0084(7)$   & $4d$ & $1$       \\
O4  & $0.5$          & $-0.0380(14)$   & $0.2954(3)$  & $0.0145(11)$  & $4e$ & $1$       \\
O5  & $0$            & $-0.0487(9)$    & $0$          & $0.0088(10)$  & $2a$ & $1$       \\
O6  & $0.5$          & $0.0542(13)$    & $0.5$        & $0.0108(10)$  & $2b$ & $1$       \\

\end{longtable}
\FloatBarrier

\begin{longtable}{c c c c c c c}
\caption{\textbf{Anisotropic displacement parameters $U_{ij}$ (\AA$^{2}$) for \LNO{} and \LPNO{}}.}
\label{SI:ref_thermal} \\

\hline
Atom & $U_{11}$ & $U_{22}$ & $U_{33}$ & $U_{12}$ & $U_{13}$ & $U_{23}$ \\
\hline
\endfirsthead

\multicolumn{7}{c}%
{{\tablename\ \thetable{} -- continued from previous page}} \\

\hline
Atom & $U_{11}$ & $U_{22}$ & $U_{33}$ & $U_{12}$ & $U_{13}$ & $U_{23}$ \\
\hline
\endhead

\hline
\multicolumn{7}{r}{{Continued on next page}} \\
\endfoot

\hline
\endlastfoot

\multicolumn{7}{c}{\LNO{}, $Am2m$ (50~K)} \\
\hline

La1 & $0.001371(9)$  & $0.001305(13)$ & $0.001345(7)$  & $0$           & $0$          & $-0.000033(5)$  \\
La2 & $0.001289(9)$  & $0.001146(13)$ & $0.001348(7)$  & $0$           & $0$          & $0.000053(6)$   \\
La3 & $0.001421(12)$ & $0.001395(14)$ & $0.001663(9)$  & $0$           & $0$          & $0$             \\
La4 & $0.001515(12)$ & $0.001451(13)$ & $0.001644(9)$  & $0$           & $0$          & $0$             \\
Ni1 & $0.001216(18)$ & $0.00125(3)$   & $0.001924(13)$ & $0$           & $0$          & $0.000042(14)$  \\
Ni2 & $0.000982(17)$ & $0.00088(2)$   & $0.001838(13)$ & $0$           & $0$          & $-0.000066(14)$ \\
O1  & $0.00295(5)$   & $0.00311(7)$   & $0.00506(6)$   & $-0.00112(5)$ & $0.00000(6)$ & $0.00010(6)$    \\
O2  & $0.00295(5)$   & $0.00302(6)$   & $0.00534(5)$   & $-0.00094(5)$ & $0.00007(7)$ & $0.00013(8)$    \\
O3  & $0.00448(11)$  & $0.00437(13)$  & $0.00322(8)$   & $0$           & $0$          & $0.00040(9)$    \\
O4  & $0.00594(12)$  & $0.00480(14)$  & $0.00309(8)$   & $0$           & $0$          & $0.00062(9)$    \\
O5  & $0.00534(18)$  & $0.00438(18)$  & $0.00227(11)$  & $0$           & $0$          & $0$             \\
O6  & $0.00428(16)$  & $0.00417(16)$  & $0.00248(11)$  & $0$           & $0$          & $0$             \\

\hline
\multicolumn{7}{c}{\LNO{}, $Amam$ (670~K)} \\
\hline

La1 & $0.01228(6)$  & $0.01320(6)$  & $0.00928(3)$ & $0$          & $0$ & $-0.000402(19)$ \\
La2 & $0.01657(13)$ & $0.01634(13)$ & $0.01293(5)$ & $0$          & $0$ & $0$             \\
Ni1 & $0.00797(10)$ & $0.00601(8)$  & $0.01075(4)$ & $0$          & $0$ & $0.00026(3)$    \\
O1  & $0.0301(10)$  & $0.0249(6)$   & $0.0125(4)$  & $0$          & $0$ & $0.0014(4)$     \\
O2  & $0.0128(5)$   & $0.0170(5)$   & $0.0242(6)$  & $0.0065(3)$  & $0$ & $0$             \\
O3  & $0.0130(5)$   & $0.0165(5)$   & $0.0272(6)$  & $-0.0063(3)$ & $0$ & $0$             \\
O4  & $0.0285(11)$  & $0.0248(8)$   & $0.0082(4)$  & $0$          & $0$ & $0$             \\

\hline
\multicolumn{7}{c}{\LPNO{}, $Am2m$ ($300\,$K)} \\
\hline

La1/Pr1 & $0.00569(15)$ & $0.00477(11)$ & $0.00439(9)$  & $0$          & $0$           & $-0.00018(7)$  \\
La2/Pr2 & $0.00595(15)$ & $0.00658(14)$ & $0.00447(9)$  & $0$          & $0$           & $0.00048(8)$   \\
La3/Pr3 & $0.0057(2)$   & $0.00601(18)$ & $0.00874(14)$ & $0$          & $0$           & $0$             \\
La4/Pr4 & $0.0077(2)$   & $0.00813(19)$ & $0.00378(11)$ & $0$          & $0$           & $0$             \\
Ni1     & $0.0040(3)$   & $0.0025(2)$   & $0.0052(2)$   & $0$          & $0$           & $0.00050(16)$  \\
Ni2     & $0.0023(3)$   & $0.0039(3)$   & $0.0053(2)$   & $0$          & $0$           & $0.00009(17)$  \\
O1      & $0.0076(4)$   & $0.0092(3)$   & $0.0126(4)$   & $-0.0020(2)$ & $-0.0024(10)$ & $0.0043(8)$    \\
O2      & $0.0072(3)$   & $0.0082(3)$   & $0.0156(4)$   & $-0.0034(3)$ & $0.0012(12)$  & $0.0031(10)$   \\
O3      & $0.0127(18)$  & $0.0078(9)$   & $0.0047(10)$  & $0$          & $0$           & $0.0007(8)$    \\
O4      & $0.016(2)$    & $0.0172(16)$  & $0.0106(15)$  & $0$          & $0$           & $0.0042(13)$   \\
O5      & $0.017(2)$    & $0.0055(10)$  & $0.0044(14)$  & $0$          & $0$           & $0$             \\
O6      & $0.0075(16)$  & $0.0193(19)$  & $0.0057(16)$  & $0$          & $0$           & $0$             \\

\end{longtable}

\FloatBarrier

%======================%%
% SI SECTION 
%====================%%
\section{\NoCaseChange{Exclusion of a monoclinic model}}
\label{sec:monocli}

In the $Amam$ structure, there exists a $2_1$ screw along the $a$ axis. This is visible from the combined operation of the $a$ glide and the mirror perpendicular to the $c$ axis:
\begin{equation}
    (x, y, z) \to (x+1/2, -y, z)\to (x+1/2, -y, -z)
\end{equation}
The $2_1$ screw imposes a reflection condition; $h\,\,0\,0$ with $h=\,$odd should be absent. This means that this reflection condition can be preserved, even if the $a$-glide symmetry is broken while the $2_1$ screw remains.
As for the $Am2m$ structure, this reflection condition is violated through the breaking of the $a$-glide plane. Indeed, we observe the $3\,0\,0$ reflection and track its temperature dependence as an order parameter of the charge order (Fig.~\ref{main:Fig3}\textbf{c}).

In previous research, a monoclinic $P2_1/m$ structure is inferred from in-house XRD experiments~\cite{Li2026-go}. The $P2_1/m$ structure, however, retains the same screw axis along the $a$-axis of $Amam$ and also the associated extinction condition on $h\,\,0\,0$, as its cell definition is $a_\mathrm{m} = -b_\mathrm{o},\ b_\mathrm{m} = a_\mathrm{o},\ c_\mathrm{m} = 0.5\cdot(-b_\mathrm{o} + c_\mathrm{o})$. 
The violation of extinction conditions for $h\,\,0\,0$ in our data excludes the $P2_1/m$ structure.

Our conclusion regarding $P2_1/m$ is further supported by the observed domain structure. Within experimental resolution, only two domains are resolved (Supplementary Fig.~\ref{SI:domain}), corresponding to orthorhombic twins. For a monoclinic structure, four domains are expected in the diffraction pattern.

%======================%%
% SI SECTION 
%====================%%
\section{\NoCaseChange{Density-wave transition observed in resistance}}
\label{sec:SI_resistivity}

\begin{figure}[ht]%
\centering
\includegraphics[width=0.5\textwidth]{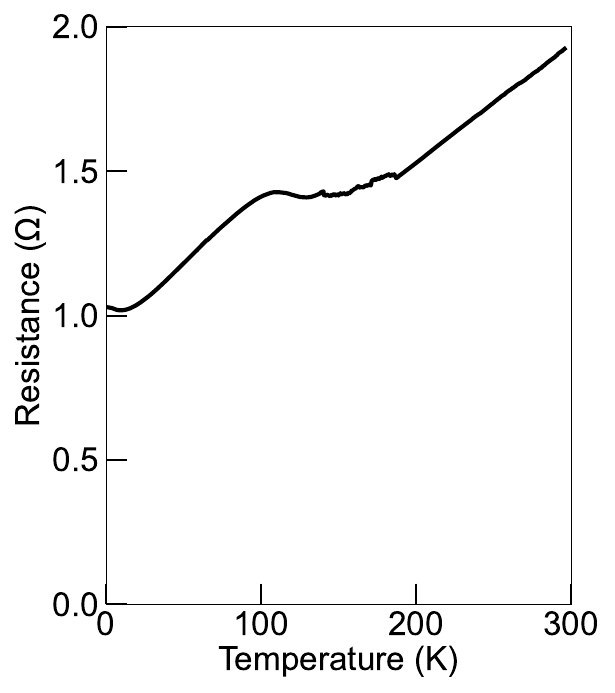}
\caption{\textbf{Density-wave transition in \LNO{}.} Temperature dependence of resistance, measured in a diamond anvil cell with a solid pressure-transmitting medium on a single crystalline piece of \LNO{} (P3). The applied pressure is $4\,$GPa, well below the critical pressure for superconductivity $\sim 10\,$GPa.
}
\label{SI:resistivity}
\end{figure}

To confirm the presence of a density-wave transition, we perform a resistance measurement on a single-crystalline piece of \LNO{} (piece P$3$) from the same batch as the one used for synchrotron XRD measurements. The sample was placed in a BeCu-type diamond anvil cell (DAC) with a $300\,\mu$m flat tip. KBr was used as the solid pressure-transmitting medium. The size of the sample was about $30\times 30 \times 15\,\mu\mathrm{m}^3$. We then performed the standard four-probe method to measure resistance. Under this condition, we only need to put the electrodes on the flat of the diamond. This helps us ensure good contact resistance to the sample. We apply a moderate pressure of $5\,$GPa, where the density-wave transition is still present~\cite{Liu2026-vk}. Supplementary Figure~\ref{SI:resistivity}\textbf{a} illustrates the temperature dependence of the resistance. The data show a clear density-wave-like anomaly at $120\sim 150\,$K, consistent with previous reports.

\FloatBarrier

%======================%%
% SI SECTION 
%====================%%
\section{\NoCaseChange{Synchrotron X-ray diffraction on \LPNO{}}}
\label{sec:LPNO}
We performed synchrotron XRD experiments using two beamlines at SPring-8, BL02B1 (bending magnet) and BL05XU (undulator). Hereafter, we refer to the data from the beamlines BL02B1 and BL05XU as Data2 and Data5, respectively. Unlike in \LNO{}, we did not observe $h\,0\,l$ ($h=\,$odd, $l=\,$even) reflections in Data2. However, the X-ray flux in BL05XU is approximately two to three orders of magnitude higher than that in BL02B1. This allows for detection of weak charge order in \LPNO{}. Three different, single-crystalline pieces of \LPNO{} were used, but they are all from the same batch: P$4$ (temperature dependence in Fig.~\ref{main:Fig3}\textbf{c}), P$5$ (XRD map in Supplementary Fig.~\ref{SI:Pr}\textbf{a}), and P$6$ (structure refinement in Supplementary Fig.~\ref{SI:Pr}\textbf{b}).

\bigskip
\textbf{Observation of $a$-glide symmetry breaking}

Supplementary Fig.~\ref{SI:Pr}\textbf{a} illustrates the $h\,0\,l$ plane from Data5. We observe $h\,0\,l$ ($h=\,$odd, $l=\,$even) reflections, indicating the broken $a$-glide symmetry. The intensity is five orders of magnitude weaker than the main Bragg reflections, consistent with their absence in Data2. Specifically, the observed intensity at $13\,0\,0$ is $4\times10^{-5}$ of that of the strongest Bragg reflection, $2\,\overline{2}\,0$, after correcting for the different attenuator transmissions. The strongest Bragg reflection was measured using a strong attenuator to avoid counting losses, whereas the intensity at $13\,0\,0$ was measured using a weak attenuator to enhance the weak signal.

\bigskip
\textbf{Constraining La/Pr occupancy}

We now move on to the structure refinement of \LPNO{}. There are two La sites in the $Amam$ structure of \LNO{}: an inner site within the bilayer and an outer site in the rock-salt layer (Fig.~\ref{main:Fig1}\textbf{a}). Upon lowering symmetry to $Am2m$, each of these two sites splits into two, yielding a total of four La sites (two inner sites, two outer sites). Now considering \LPNO{} in space group $Am2m$, we constrain the two inner sites (two outer sites) to have the same Pr occupancy, as justified in the following.

For the sake of argument, assume the La/Pr occupancy differed between the two inner sites or between the two outer sites in $Am2m$. First, this implies that the La/Pr occupancy must change across the charge-ordering transition at $T_\mathrm{CO}$. It is unlikely that the charge-ordering transition within the Ni-O bilayer should alter the La/Pr site occupancies. Second, if it were so, we expect strong $h\,0\,l$ ($h=\,$odd, $l=\,$even) intensity due to the sizable difference in the scattering factors of La and Pr. The weakness of these reflections indicates that the site ordering of La/Pr, and concomitant occupational differences, are negligible within a layer (inner or outer layer). We therefore constrain the two sites in each layer to have the same La/Pr occupancy.

\bigskip
\textbf{High-angle analysis}

The observation of $h\,0\,l$ ($h=\,$odd, $l=\,$even) reflections in Data5 is achieved by saturating strong Bragg intensities. This results in a significant counting loss for the strong reflections and makes it challenging to perform quantitative structure refinement incorporating both weak and strong reflections. Therefore, we perform structure refinement using Data2, as we also did for \LNO{}. The $a$-glide symmetry breaking affects the structure factor at most reflections, including those other than the $h\,0\,l$ ($h=\,$odd, $l=\,$even) reflections forbidden in $Amam$. Therefore, the refinement with the $Am2m$ model is decisive even when reflections at $h\,0\,l$ ($h=\,$odd, $l=\,$even) are not observed in Data2.

In contrast to the single-domain data for \LNO{}, Data2 for \LPNO{} include contributions from two orthorhombic twins. We find that the peak splitting at low angles is too small to be fully resolved. Restricting the refinement to high-angle reflections avoids uncertainties arising from these overlaps. We thus performed high-angle analysis for \LPNO{}, using reflections from Data2 with $\sin\theta/\lambda > 0.6$. After determining the structural parameters, we evaluated the refinement statistics using all reflections, except the overlapping low-angle reflections. The refinement includes four domains from orthorhombic and inversion twinning (Supplementary Table~\ref{Tab:LPNO_twin}). The inversion-related domains were incorporated into the HKLF~5 reflection file using HKLF5Tools~\cite{Ivlev2019-el}.

\bigskip
\textbf{Result of structure refinement}

We present the results of the structure refinement for the $Am2m$ model in \LPNO{} at room temperature in Supplementary Tables~\ref{Tab:SI_refinement}-\ref{Tab:LPNO_twin}.
The model with relaxed oxygen occupancy does not improve the agreement with experiment compared to the model with stoichiometric oxygen occupancy.
As in \LNO{} (Table~\ref{Tab:refinement}), our data are consistent with ideal oxygen stoichiometry in \LPNO{}. Supplementary Fig.~\ref{SI:Pr}\textbf{b} illustrates that praseodymium atoms prefer to occupy the outer rock-salt layer [occ.(Pr1)=occ.(Pr2)=$0.36(3)$ (site multiplicity 4) and occ.(Pr3)=occ.(Pr4)=$0.11(3)$ (site multiplicity 2)], consistent with a previous study using electron energy loss spectroscopy~\cite{Dong2025-ni}. The total site occupancy is estimated to be La:Pr$=2.17(5): 0.83(5)$. The bond-valence sum (BVS) is calculated as $+2.62(2)$ and $+2.47(2)$ for Ni1 and Ni2, respectively. The difference in the BVS is $0.14(3)$, which is about three times smaller than that in \LNO{}, $0.478(2)$. The smaller BVS difference agrees with the weaker intensities at $h\,0\,l$ ($h=\,$odd, $l=\,$even) in \LPNO{}.

\begin{figure}[ht]%
\centering
\includegraphics[width=0.8\textwidth]{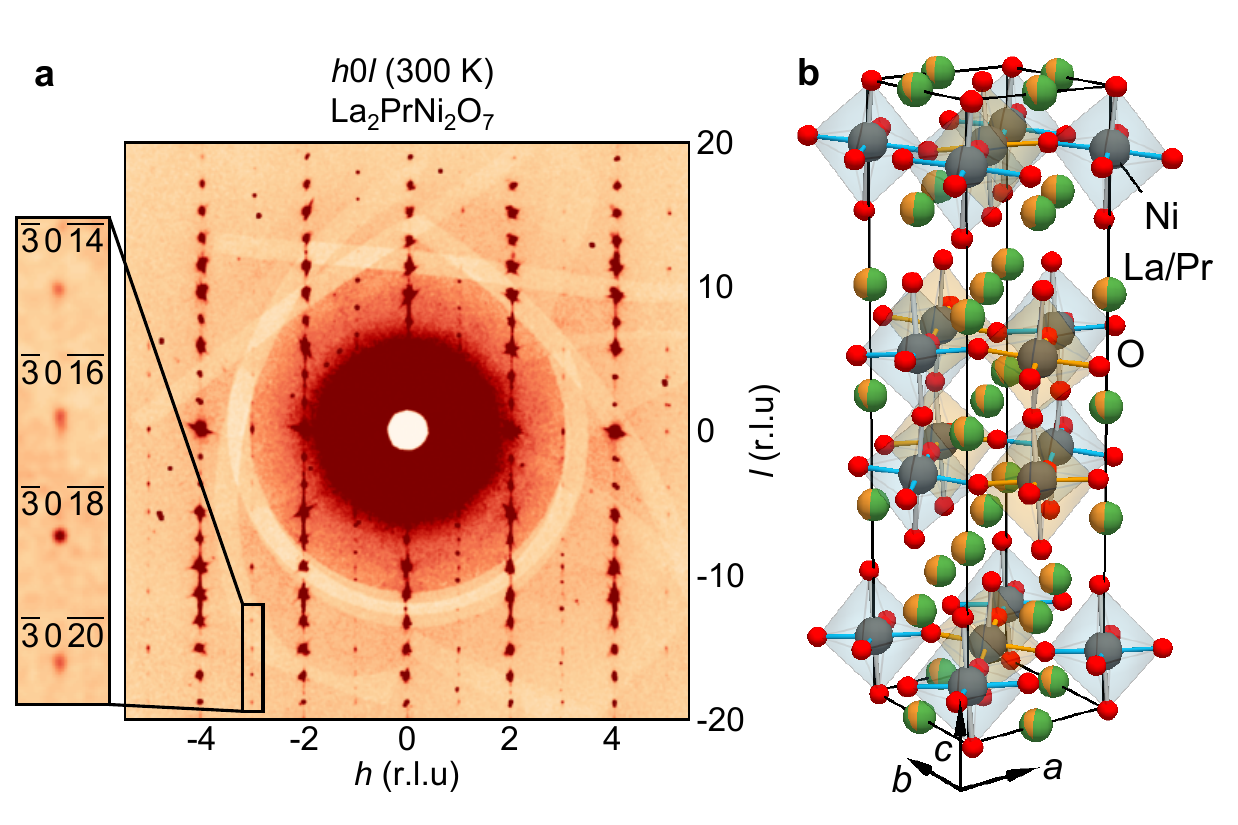}
\caption{\textbf{Charge order in \LPNO{}.} \textbf{a}, X-ray diffraction pattern on the $h\,0\,l$ plane at $300\,$K. Here, $-3\,0\,l$ reflections associated with the broken $a$-glide symmetry are emphasized by the black rectangle. \textbf{b}, Polar orthorhombic $Am2m$ structure for \LPNO{}. Green, orange, gray, and red balls indicate lanthanum, praseodymium, nickel, and oxygen atoms, respectively. Praseodymium prefers the outer rock-salt layer.
}
\label{SI:Pr}
\end{figure}

\begin{table}[htbp]
\centering
\renewcommand{\arraystretch}{1.2}
\setlength{\tabcolsep}{10pt}
\caption{\textbf{Twin matrix for $Am2m$ model of \LPNO{} at 300\,K}.}
\label{Tab:LPNO_twin}
\begin{tabular}{ccc}
\hline
Domain & Twin matrix & Fraction \\
\hline
1 &
$(1,0,0;\,0,1,0;\,0,0,1)$ &
$0.40(20)$ \\

2 &
$(0,-1,0;\,1,0,0;\,0,0,1)$ &
$0.44(12)$ \\

3 &
$(-1,0,0;\,0,-1,0;\,0,0,-1)$ &
$0.13(13)$ \\

4 &
$(0,1,0;\,-1,0,0;\,0,0,-1)$ &
$0.07(12)$ \\
\hline
\end{tabular}
\end{table}

\FloatBarrier

%======================%%
% SI SECTION 
%====================%%
\section{\NoCaseChange{Ligand-hole state and Zhang-Rice singlet formation}}
\label{sec:ZRS}

We discuss a potential microscopic picture of the charge order in \LNO{}, based on the state-of-the-art understanding of charge and spin ordering in perovskite \textit{R}NiO$_3$ (\textit{R} = rare-earth), and connect the charge order in \LNO{} to spin density wave (SDW) order.

\bigskip
\textbf{Charge order via ligand hole in $\bm{R}$NiO$_3$}

The perovskite family \textit{R}NiO$_3$ is known to exhibit a metal-insulator transition driven by charge order~\cite{Alonso1999-ld}. In this case, the formal Ni valence is integer, Ni$^{3+}$ ($d^7$), so that the term ``charge disproportionation" is frequently used. Similar to our finding in \LNO{}, one Ni site splits into two Ni sites at the transition, with two distinct in-plane Ni-O bond lengths, corresponding to checkerboard charge order.

More recently, theoretical studies of \textit{R}NiO$_3$ propose that the charge order is driven by ligand-hole formation~\cite{Johnston2014-jt,Green2016-du}. In this picture, the parent state of Ni is $d^{8}\underline{L}$ with $\underline{L}$ denoting one hole transferred to a ligand oxygen, rather than the formal $d^{7}$. At the transition point, the Ni site splits into two sites with $d^{8}$ and $d^{8}\underline{L}^2$. Since one site forms weaker (the other site stronger) covalent bonding with oxygen as compared to the parent phase, this results in an expanded (contracted) octahedral environment for Ni. In this sense, the charge order in \textit{R}NiO$_3$ is referred to as a ``bond disproportionation". Experimentally, the ligand-hole scenario in \textit{R}NiO$_3$ has been verified through resonant inelastic X-ray scattering~\cite{Lu2018-fd}.

Although the formal Ni valence is different, being Ni$^{2.5+}$ ($d^{7.5}$) in \LNO{}, the charge order reported here shares common features with perovskite \textit{R}NiO$_3$. Both charge-ordered states have similar contracted and expanded octahedral environments. The ligand-hole picture may also be consistent with recent X-ray absorption and resonant inelastic X-ray scattering experiments on \LNO{}~\cite{Chen2024-rixs,Takegami2024-fs}. 

\bigskip
\textbf{Zhang-Rice singlet formation and relevance to magnetism in \LNO{}}

Theoretical studies of \textit{R}NiO$_3$ also propose that the ligand-hole state further induces the Zhang-Rice singlet (ZRS)~\cite{Green2016-du}. For the Ni$1$ site with a $d^{8}$ configuration, there exists one hole in each orbital of the $e_g$ manifold, which contains $d_{x^2-y^2}$ and $d_{z^2}$ orbitals. In the local picture, this corresponds to a total spin of $S=1$. On the other hand, in the case of $d^{8}\underline{L}^2$, the two holes of Ni$2$ and the ligand holes form a ZRS, resulting in $S=0$. Therefore, checkerboard charge order naturally induces checkerboard spin order with distinct moment sizes on two nickel sites, as experimentally observed in \textit{R}NiO$_3$~\cite{Alonso1999-ld}.

This ZRS scenario was proposed for \LNO{} in a recent theoretical work~\cite{Foyevtsova2025-fs}. The ZRS accounts for the large difference in magnetic moments on the two Ni sites in \LNO{}, as observed in recent experiments~\cite{Plokhikh2025-main,Khasanov2025-main,Chen2026-main}. In other words, the checkerboard charge order in \LNO{}, emerging below $T_\mathrm{CO} = 560\,$K, can be seen as a precursor to the spin density wave transition at a lower temperature.

\end{document}